\documentclass[fleqn,usenatbib]{mnras}

\usepackage{txfonts}

\usepackage[T1]{fontenc}

\DeclareRobustCommand{\VAN}[3]{#2}
\let\VANthebibliography\thebibliography
\def\thebibliography{\DeclareRobustCommand{\VAN}[3]{##3}\VANthebibliography}

\usepackage{graphicx}	
\usepackage{amssymb}	
\usepackage{epstopdf}
\usepackage{comment}
\usepackage{xcolor,colortbl}
\usepackage{multirow}

\usepackage{lscape} 
\usepackage{threeparttable} 

\title[Accretion Scenario of MAXI J1820+070]{Accretion Scenario of MAXI J1820+070 during 2018 Outbursts with Multi-mission Observations}

\author[Geethu et al.]{
Geethu Prabhakar$^{1}$\thanks{geethuprabhakar.17@res.iist.ac.in},
Samir Mandal$^{1}$,
Athulya M. P.$^{2}$,
and Anuj Nandi$^{3}$
\\
$^{1}$ Department of Earth and Space Sciences, Indian Institute of Space Science and Technology (IIST), Trivandrum - 695547, India\\
$^{2}$ Department of Physics, Dayananda Sagar University, Bengaluru - 560068, India\\
$^{3}$ Space Astronomy Group, ISITE Campus, U R Rao Satellite Centre, Bengaluru - 560037, India
}

\date{Accepted XXX. Received YYY; in original form ZZZ}

\pubyear{2015}

\begin{document}
\label{firstpage}
\pagerange{\pageref{firstpage}--\pageref{lastpage}}
\maketitle

\begin{abstract}
We present a comprehensive spectral and temporal study of the black hole X-ray transient MAXI J1820+070 during its outbursts in 2018 using Swift/XRT, NICER, NuSTAR and AstroSat observations. The Swift/XRT and NICER spectral study shows a plateau in the light curve with spectral softening (hardness changes from $\sim$ $2.5$ to $2$) followed by a gradual decline without spectral softening during the first outburst. Also, spectral modelling suggests that the first outburst is in the low/hard state throughout with a truncated disk whereas the thermal disk emission dominates during the second outburst.
During the entire outburst, strong reflection signature (reflection fraction varies between $\sim$ $0.38 - 3.8$) is observed in the simultaneous wideband (NICER-NuSTAR, XRT-NuSTAR, AstroSat) data due to the presence of a dynamically evolving corona. The NICER timing analysis shows Quasi-periodic Oscillation (QPO) signatures and the characteristic frequency increases (decreases) in the plateau (decline) phase with time during the first outburst. We understand that the reduction of the electron cooling timescale in the corona due to spectral softening and the resonance oscillation with the local dynamical timescale may explain the above behavior of the source during the outburst. Also, we propose a possible scenario of outburst triggering and the associated accretion geometry of the source.
\end{abstract}

\begin{keywords}
accretion, accretion discs - black hole physics - X-rays: binaries - stars: individual: MAXI J1820+070
\end{keywords}



\section{Introduction}

Accretion physics around black hole X-ray binaries (BH-XRBs) has been an exciting topic in Astronomy since the seventies. However, the process of accretion and the geometry of the accreting system remain a poorly understood area.  
BH-XRBs are systems that consist of a black hole (BH) as the primary compact object and a non-compact secondary star as the companion, both revolve around their common centre of mass. The accreted matter from the companion star eventually falls onto the BH, emitting radiations, predominantly, in X-rays.
Based on the nature of the X-ray activities, BH-XRBs are classified into persistent and transient sources \citep{2016yCat..22220015T,2016A&A...587A..61C,2018JApA...39....5S}. Persistent sources are continuously active and accreting consistently at a high rate, whereas the transient sources show repeated transitions between outbursts and quiescent states \citep{2006ARA&A..44...49R}. 

The energy spectral modelling of BH-XRBs generally needs a multicolor blackbody
component, which is assumed to be emitted from an accretion disc \citep{1973A&A....24..337S} and a non-thermal \textit{powerlaw} component resulted due to the Comptonization \citep{1980A&A....86..121S,1994MNRAS.269L..55Z,1985A&A...143..374S,1994AIPC..304..380T} of the thermal photons from the accretion disc by a `hot' electron cloud known as corona \citep{1995ApJ...455..623C,1995ApJ...452..710N,2015ApJ...807..108I}.
The observed X-ray spectrum shows additional components if a fraction of the hard photons emitted from the corona illuminates the accretion disk. The hard X-ray photons are reprocessed (absorbed/reflected) in the optically thick accretion disk and gives rise to the reflection component \citep{1988ApJ...335...57L, 2005MNRAS.358..211R}. The Fe $K_\alpha$ emission line at 6.4 keV is one of the important components in the reflection spectrum. It is often relativistically broadened, having a gaussian width of several hundred eV, and possesses an asymmetric line profile \citep{1993ARA&A..31..717M}. Other important features that constitute the reflection spectra are a soft excess at $\sim 1$ keV \citep{2016MNRAS.462..751G,2018cosp...42E3408T} and a Compton hump between $20-50$ keV \citep{2000PASP..112.1145F,2003PhR...377..389R,2005MNRAS.358..211R,2015MNRAS.454.4440W,2021SSRv..217...65B}. These features are clearly observed with the arrival of good spectral resolution solid-state instruments like the \textit{Nuclear Spectroscopic Telescope Array} (NuSTAR), which are capable of focusing hard X-rays ($\gtrsim$ 10 keV) \citep{2013ApJ...770..103H}.

In addition to spectral variability, these sources also possess variability in their light curves. The power density spectra (PDS) show broadband noise features \citep{1989ARA&A..27..517V} and features with concentrated power at a characteristic frequency. QPOs are fundamentally classified into two categories, high-frequency ($30 - 10^{3}$ Hz) \citep{1997ApJ...482..993M,2002ApJ...580.1030R}  and low-frequency ($10^{-2}$ - $30$ Hz) \citep{2002ApJ...564..962R} QPOs. The low-frequency QPOs (LFQPO) are identified as type-A, B, and C \citep{2002ApJ...564..962R,2005ApJ...629..403C}.

Based on the spectral and temporal variabilities, accretion states of the BH-XRBs are categorized as Low/Hard State (LHS), High/Soft State (HSS) and Intermediate States (Hard Intermediate State (HIMS) and Soft Intermediate State (SIMS)) \citep{2001ApJS..132..377H,2005Ap&SS.300..107H,2006ARA&A..44...49R,2012A&A...542A..56N,2014AdSpR..54.1678R,2019MNRAS.487..928S,2021MNRAS.502.1334S}. 
In the LHS, the initial stage of the outburst, the emission is mainly non-thermal and has a spectrum approximated by a \textit{powerlaw} with a high-energy cut-off. The source is faint, and the radiation spectrum has a photon index $\Gamma \sim 1.7$. The PDS at this state show high timing variability (root mean square (RMS) variability $\gtrsim 20\%$) and type-C QPOs. This state is often associated with quasi-steady radio jets.
In the HSS, the spectrum is dominated by thermal emission from the accretion disc. The source appears brighter in this state, and the spectrum has a photon index $\Gamma \sim 3$. Timing variability is low i.e., PDS are almost featureless; QPOs are absent or very weak.
Intermediate states lie between LHS and HSS states and are relatively short-lived. HIMS state is associated with type-C QPOs, whereas SIMS possess type-A or/and type-B QPOs and are less common \citep{2005A&A...440..207B}.

Transient sources spend most of their time in quiescent state, having a very low luminosity and a non-thermal hard spectrum. 
The nature of outburst varies between the BH systems; even different outbursts of the same system show different evolution trends \citep{2019MNRAS.486.2705A,2020A&A...637A..47A}. However, the behavior of the outbursts in a \textit{hardness-intensity diagram} (HID), which is a plot of the X-ray intensity versus hardness ratio (HR), shows some general regulations \citep{2001ApJS..132..377H,2004MNRAS.355.1105F,2005AIPC..797..197B,2006ARA&A..44...49R,2012A&A...542A..56N,2012MNRAS.427..595M,2014AdSpR..54.1678R}.
A typical HID forms a q-shaped diagram where the source evolves over the four spectral states (LHS, HIMS, SIMS and HSS) in a counterclockwise direction. 
An outburst is termed as `complete/successful' if the source moves from the quiescent state to the LHS and the luminosity keeps on increasing till the source reaches the HSS via intermediate states. The source never goes to HSS for a `failed' outburst. 

Similarly, the global timing behavior of transients is expressed in the \textit{RMS-intensity diagram} (RID) to track the evolution of the intensity and RMS. The \textit{hardness-RMS diagram} (HRD) \citep{2010LNP...794...53B,2012MNRAS.427..595M,2016MNRAS.462.1834R}, where the fractional RMS is plotted against the HR, is used to understand the spectro-temporal variations during the outbursts.

MAXI J1820+070 (ASASSN-18ey) is a galactic low mass X-ray binary (LMXB) discovered in optical by All-Sky Automated Survey for Super Novae (ASAS-SN) \citep{2014ApJ...788...48S} on 3 March 2018. A week later, on 11 March 2018, it was discovered as an X-ray transient by MAXI/GCS \citep{2018ATel11399....1K, 2018ATel11400....1D, 2018ApJ...867L...9T}. It is located at $RA=18^h 20^m 21^s.9$, $DEC=+07^\circ 11^{'} 07^{''}.3$
 (J2000) \citep{2018ATel11399....1K}. It is one of the brightest X-ray transients ever discovered, having a soft X-ray flux of $\sim$ 4 Crab in $2-6$ keV \citep{2019ApJ...874..183S} and has a low column density of $1.5 \times 10^{21} \rm cm^{-2}$ \citep{2018ATel11423....1U}. The source is observed in radio  \citep{2018ATel12061....1B} also. There are many follow-up observations done by different X-ray missions such as Swift, NICER, NuSTAR, HXMT, etc.

Studies on the optical counterpart of MAXI J1820+070 by \citet{2019ApJ...882L..21T} derived a mass function $f(M) =5.18 \pm 0.15 M_\odot$, which is the dynamical confirmation of the black hole nature of the source and estimated its mass as $7-8 M_\odot$. The distance to the source is obtained as $2.96 \pm 0.33$ kpc by radio parallax method using VLBA \citep{2020MNRAS.493L..81A}. The VLBI observation provides an inclination angle ($\theta$) of $63 \pm 3^0$ and mass of $9.2 \pm 1.3 M_\odot$ \citep{2020MNRAS.493L..81A}. Later, \citet{2020ApJ...893L..37T} did a more accurate calculation of the mass ratio $ q = 0.072 \pm 0.012$, which leads to $M = 8.48^{+0.79}_{-0.72} M_\odot$ using $\theta=63 \pm 3^0$ \citep{2020MNRAS.493L..81A}. \citet{2019MNRAS.490.1350B} estimated the inner radius of the accretion disc using NuSTAR observations in the hard state and suggest a low spinning BH.
\citet{2021MNRAS.504.2168G} reported a spin of $0.2^{+0.2}_{-0.3}$ by continuum fitting method in the soft state data of MAXI J1820+070 using Insight-HXMT observations.
Recently, \citet{2021MNRAS.508.3104B} carried out the spin measurement using Relativistic Precession Model (RPM) on NICER observations and the estimated value of the spin as $0.799^{+0.016}_{-0.015}$ under the assumption of a BH mass of $8.48 M_\odot$. 

The spectral and temporal studies in the LHS with NICER data reveal the corona as a contracting lamp-post and the inner disk is having non-truncating nature \citep{2019Natur.565..198K}. \citet{2019MNRAS.490.1350B} also suggests a steady inner accretion disk in the LHS and found that the variability in the spectrum is produced by the variation in the coronal geometry. Such an evolving corona has been reported in BH-XRBs \citep{2006A&A...448.1125M} before as well.
\citet{2021ApJ...909L...9Z,2021MNRAS.507.2744A,2021A&A...654A..14D} challenged this model, and suggested a truncated disc in the LHS of MAXI J1820+070. Type-C QPOs are detected in the LHS of the source \citep{2018ATel11488....1M,2018ATel11510....1Y,2018ATel11723....1Z,2018ATel11824....1F,2021NatAs...5...94M}. \citet{2020ApJ...891L..29H} has detected a type-B QPO during the hard to soft state transition.

There are many studies \citep{2018ApJ...868...54S,2019MNRAS.490.1350B,2019Natur.565..198K,2019ApJ...879L...4M,2019MNRAS.490L..62P,2020ApJ...889..142S,
2020ApJ...889L..17M,2020ApJ...896...33W,2021MNRAS.504.3862T,2021MNRAS.508.3104B,2021ApJ...910L...3W, 2021MNRAS.505.3452P,2021arXiv211208116Z,2021MNRAS.508.3104B,2021MNRAS.506.2020D,2021NatAs...5...94M,
2021ApJ...910...21R,2021A&A...656A..63M,2021MNRAS.507...55M,2022MNRAS.tmp..101K} that address various spectral and temporal properties of the source. But these works mainly focus on the LHS of the source.
A complete comprehensive spectral study of the entire outburst, particularly in the HSS, has not been done. The purpose of this manuscript is to understand the 2018 double outbursts of the source in total using multi-instrument spectral and timing properties. The reflection modelling of NuSTAR data alone has performed for selected observations before but never throughout the outbursts. Here, we considered the entire 2018 outburst for reflection studies and extend it to lower energy band using NICER-NuSTAR, XRT-NuSTAR, AstroSat and attempt to understand the evolution of the corona. 

We study the spectral and timing properties of MAXI J1820+070 using multiple instruments: Swift/XRT, NICER, NuSTAR, and AstroSat covering the entire 2018 outburst. We present the observation and data reduction of various instruments in \S\ref{sec:redn}. The spectral analysis and modelling of Swift/XRT and NICER is done in \S\ref{sec:xrt-ni}. We discuss the HID and the evolution of the spectral parameters in \S\ref{sec:lc-hid} and \S\ref{sec:evol}, respectively. The simultaneous wideband spectral study is discussed in \S\ref{sec:simu}. The timing studies are presented in \S\ref{sec:tempo}. We present a possible accretion scenario during the entire outburst in \S\ref{sec:accsin}, and finally, we conclude in \S\ref{sec:concl}.

\section{OBSERVATIONS AND DATA REDUCTION}
\label{sec:redn}
We carry out the present study using four different instrument data, i.e., Swift/XRT, NICER, NuSTAR, and AstroSat over a period from 12 March 2018 to 21 November 2018. During this period, the source has undergone two outbursts: the first one (Outburst-I) is a `failed' outburst followed by a `successful' one (Outburst-II). 
An observation summary for various instruments is given in Table~\ref{tab:obs}. We analyse and model all the data from Swift/XRT, NICER and AstroSat observations in this period and consider NuSTAR only for simultaneous observations with NICER and XRT.

\subsection{Swift/XRT Data Reduction}
\label{sec:swxrt}
Swift X-ray telescope (XRT) operates in the energy range $0.3-10$ keV.
MAXI J1820+070 is a very bright source, and XRT observed the source in windowed timing (WT) mode. It covers a field of view of 8 arcmin and provides one-dimensional imaging information. The time resolution of this mode is $1.7$ ms. We use XRT data of the source from MJD 58189.02 to MJD 58434.35 (Table~\ref{tab:obs}). 
Data reduction is carried out using standard procedures according to the instrument website\footnote{\url{https://www.swift.ac.uk/analysis/xrt/}}. Level 2 products are generated from \textit{xrtpipeline} script. We select a circle of radius 30 pixels around the image centre as the source region, and an annular region of $70-130$ pixel radii is taken as the background region.
If the source flux is too high, the observed data become piled-up. The pile-up limit for WT mode data is $\sim$ 100 counts sec$^{-1}$. For such data, we select an annular source region of outer radius 30 pixel and varying inner radius for the analysis. The size of the inner radius is determined such that the data in the annular region remain pile-up free.  
XSELECT (V2.4g) is used to produce standard light curve and spectra corresponding to the source and background regions. Then, we edit the \texttt{BACKSCAL} keyword in the header files of the source and background spectrum for each observation according to the size of the corresponding regions selected. The ancillary response function (ARF) files are generated using the \texttt{xrtmkarf} tool with the source spectra and exposure map. We group the source spectra into 20 photons per bin, and no systematics is added. We use the response files (RMF) from the CALDB version 20190910\footnote{\url{https://heasarc.gsfc.nasa.gov/docs/heasarc/caldb/data/swift/xrt/index/cif_swift_xrt_20190910.html}}.

\subsection{NICER Data Reduction} 
\label{sec:nicer}
NICER's X-ray Timing Instrument (XTI) consists of an aligned collection of 56 X-ray concentrator optics (XRC) having large effective area (1900 cm$^2$ each) and silicon drift detector (SDD) pairs which detect individual photons, records their energies and times of arrival of photons with good resolution. 
It operates in the energy range of $0.2-12$ keV. NICER had good coverage over all the phases of MAXI J1820+070 since its discovery. We analyse NICER data of the source between MJD 58189.58 and  MJD 58443.82 (Table~\ref{tab:obs}). The data reduction is done using the \texttt{NICERDAS}\footnote{\url{https://heasarc.gsfc.nasa.gov/docs/nicer/nicer_analysis.html}} tools in HEASOFT v6.26.1 with the 20200722 \footnote{The results are not affected by this version of HEASOFT and NICER caldb. We verified that the same conclusion can be drawn with the currently recommended NICER calibration.} caldb version. Out of the 56 focal plane modules (FPMs) of NICER/XTI, FPM-11, 20, 22, and 60 are not functional. In addition to that, we exclude FPM-14 and 34 due to their increased noise level. During the brightest phase of the source (Outburst-II), most of the detectors were turned off to avoid telemetry saturation. We checked the total number of active detectors in each observation and take account of the reduction in the effective area. All observations are processed with \texttt{nicerl2} task and applied barycenter corrections using \texttt{barycorr} and used \texttt{refframe="ICRS"}.
The resulting event files are used to generate spectra and light curves with a time bin of 0.003 sec using \texttt{XSELECT (V2.4g)}. The ARF and RMF files are computed for each observation based on the number of active detectors. The background files are generated using \texttt{nibackgen3C50}\footnote{\url{https://heasarc.gsfc.nasa.gov/docs/nicer/tools/nicer_bkg_est_tools.html}}  \citep{2021arXiv210509901R}. We grouped the source spectra into 25 photons per bin and added a systematics of 1.5 $\%$ \citep{2020MNRAS.493.5389F}. 
For timing analysis, we generated the light curves of NICER observations with exposure time greater than 1 ksec, with a binsize of 0.003 sec for the full energy range.

\begin{table}
	\centering
	\caption{Summary of Observations of MAXI J1820+070}
	\label{tab:obs}
	\begin{tabular}{c|c|c} 
		\hline
		Mission & Start date (MJD) & End date (MJD)\\
		\hline
		&& \\
		Swift/XRT$^\dagger$ & 12-03-2018 (58189.02) & 12-11-2018 (58434.35) \\
		NICER  &   12-03-2018 (58189.58) & 21-11-2018 (58443.82) \\
		NuSTAR$\dagger \dagger$ & 14-03-2018 (58191.90) & 29-10-2018 (58420.00) \\
		&& \\ 
		\hline
		& \multicolumn{2}{c}{31-03-2018 (58208)} \\
	 	
		AstroSat$\ddagger$ & \multicolumn{2}{c}{10-07-2018 (58309)}  \\
 
		& \multicolumn{2}{c}{25-08-2018 (58355)} \\
		\hline

	\end{tabular}
        \begin{tablenotes}
    \item[a] $^\dagger$ Swift/XRT observations from MJD 58298 to MJD 58386 are ignored due to huge  pile-up.
	\item[b] $\dagger \dagger$ Only the NuSTAR observations simultaneous to NICER and XRT during this period, are considered.
	\item[c] $\ddagger$ There are only three AstroSat observations throughout our study.
	
\end{tablenotes}  
\end{table}

\subsection{NuSTAR Data Reduction}
\label{sec:nustar} 
NuSTAR is the first X-ray telescope that uses focusing optics on detecting hard X-rays. It is suitable for observing bright sources without pile-up issues. There are two focal plane module telescopes (FPMA and FPMB) in NuSTAR and operate in the energy range $3-78$ keV. We have considered only the NuSTAR observations, simultaneous with NICER and XRT between MJD 58191.9 and MJD 58420 (Table~\ref{tab:obs}). Data reduction is done using NuSTARDAS\footnote{\url{https://heasarc.gsfc.nasa.gov/docs/nustar/analysis/}} software and CALDB version 20191219. We used the \texttt{nupipeline} \texttt{(v0.4.6)} task for filtering event files with keyword \texttt{saamode} set to \texttt{strict}, \texttt{tentacle} to \texttt{yes} and \texttt{statusexpr="STATUS==b0000xxx00xxxx000"} \citep{2019MNRAS.490.1350B}. We used a circular region of radius 30 pixels centered on the brightest pixel to extract the science products for both telescopes, FPMA and FPMB. Background spectrum is extracted from another 30 pixel circular region away from the source position. The \texttt{NUPRODUCTS} task is used to generate science products such as light curves, energy spectra, response matrix files (RMF), and auxiliary response files (ARF) for both telescopes FPMA and FPMB. We grouped both spectra with a minimum of 50 counts per bin, and no systematics is added.

\subsection{AstroSat Data Reduction}
\label{sec:astrosat}
AstroSat \citep{2006AdSpR..38.2989A} consists of three X-ray instruments, namely, the Soft X-ray Telescope (SXT), Large Area X-ray Proportional Counters (LAXPC) and Cadmium-Zinc-Telluride Imager (CZTI).
We consider SXT and LAXPC data for this work. SXT is a focusing X-ray instrument working in the energy range $0.5-8$ keV and LAXPC operates between $3-60$ keV. There are three AstroSat observations (see Table~\ref{tab:obs}) of this source throughout our study. 
Level-2 SXT data from the latest pipeline version \textit{1.4b} is directly obtained from \texttt{ISSDC}\footnote{\url{https://www.issdc.gov.in/}}. 
All SXT observations of the source are in Photon Counting (PC) mode. The SXT instrument team\footnote{\url{https://www.tifr.res.in/~astrosat_sxt/dataanalysis.html}} provides the guidelines for the data reduction, and we followed \citet{2020MNRAS.497.1197B} and \citet{2021MNRAS.501.6123K} for further information.
The individual event files corresponding to each orbit are merged into a single cleaned event file to avoid any time-overlap in the events from consecutive orbits using Julia based \texttt{SXTevtmerger} script. \texttt{XSELECT v2.4g} is used to generate the light curve, spectrum, and image from the merged event file. The extracted SXT images for the last two observations (MJD 58309 and MJD 58355) show no counts in the central region because of high pile-up\footnote{\url{http://www.iucaa.in/~astrosat/AstroSat_handbook.pdf}}. Large pile-ups can saturate the CCD electronics, leading to the overflow of photons, thus registering zero counts in the center.
Therefore, we exclude the central dark patch through annular extraction with an inner radius of 3 arcmin and an outer radius of 16 arcmin for these two observations. Additionally, we observed pile-up in the energy spectra beyond 4 keV. We selected events with grade 0 (pure X-ray events) as suggested by \citet{2020arXiv201201800S} to solve this issue.
The final spectrum thus obtained is grouped into 30 photons per bin and used for the spectral analysis. A suitable ARF file for the selected region is generated using \texttt{sxtARFModule} provided by the SXT instrument team. Also, the background and RMF files required for spectral modelling are provided by the SXT team. 

Among the three LAXPC units onboard AstroSat, we only analyse data from \textit{LAXPC 20}\footnote{\url{http://astrosat-ssc.iucaa.in/}}. LAXPC Level-1 data are reduced following \citet{2017ApJS..231...10A} using \texttt{LaxpcSoft v3.4}\footnote{\url{http://astrosat-ssc.iucaa.in/?q=laxpcData}} released on 14 June 2021. The source spectra and light curves are extracted from Level-2 data following \citet{2018MNRAS.477.5437A} and \citet{2019MNRAS.487..928S}. The background and spectral response (RMF) are generated as mentioned in \citet{2017ApJS..231...10A}. We choose the background model made available by \texttt{LaxpcSoftv3.4} closest to a particular observation as the background model. To retain minimal residue in the spectrum beyond 30 keV, we opt for the events only from the top anode layer of \textit{LAXPC 20} and generate spectra and light curves in $3-60$ keV. For both SXT and LAXPC, we used a systematics of 2\% as prescribed by the instrument team\footnote{\url{https://www.tifr.res.in/~astrosat_sxt/dataana_up/readme_sxt_arf_data_analysis.txt}}.

\section{Modelling and Results}
\label{sec:result}
\begin{figure}
	\includegraphics[width=\columnwidth]{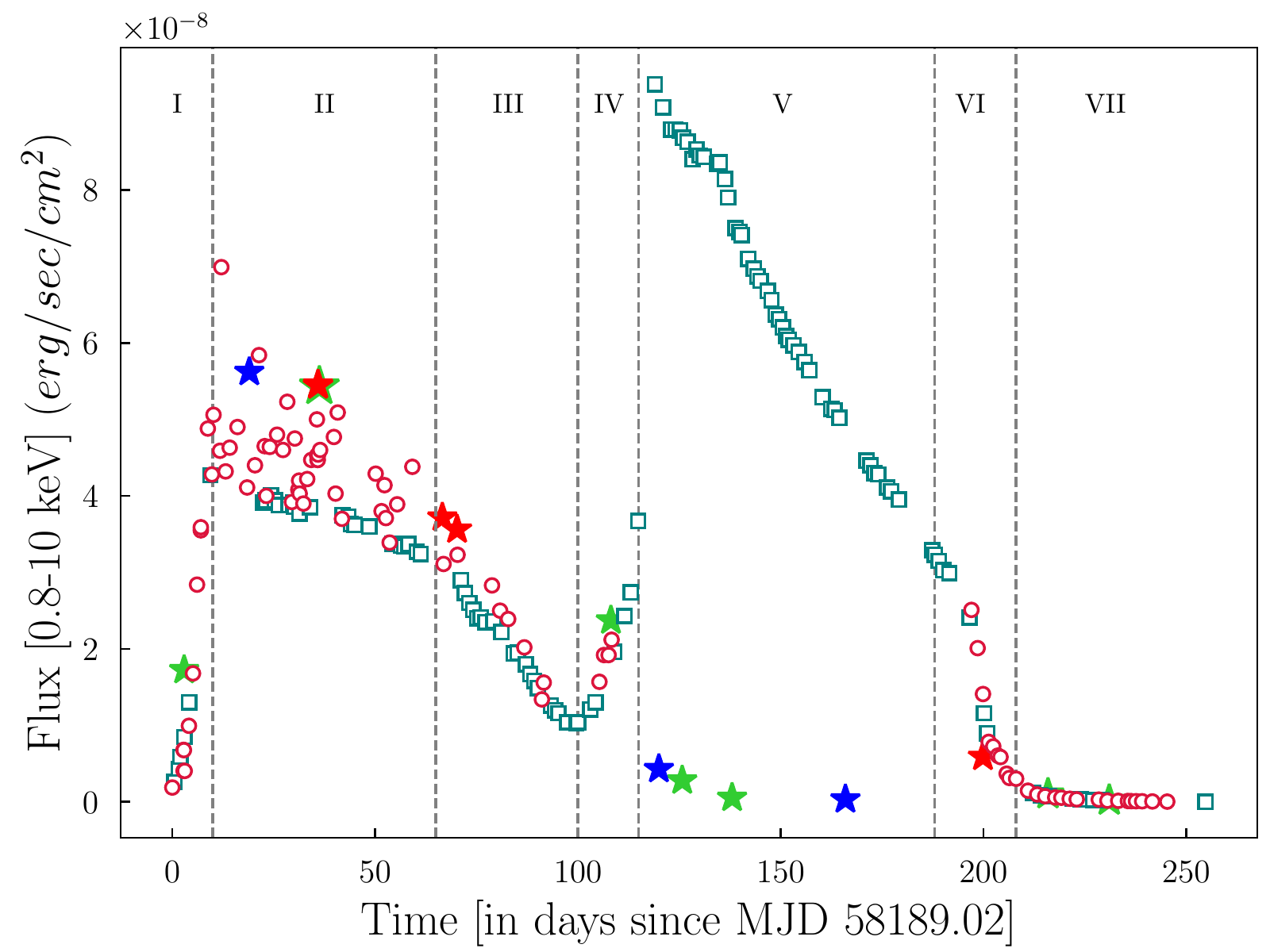}
	\caption{Light curve in the energy range $0.8-10$ keV generated using the NICER and XRT observations of MAXI J1820+070. The red circles represent the XRT data, and the green squares represent NICER data. The flux in the $10-60$ keV band for the simultaneous pairs for XRT-NuSTAR, NICER-NuSTAR and AstroSat observations are marked with red, green and blue stars respectively. Different evolutionary phases of the outbursts are marked from I to VII (see text for details).} 
	\label{fig:lc_all}
\end{figure}

MAXI J1820+070 has undergone two successive outbursts (Outburst-I \& Outburst-II) during 12 March 2018 to 21 November 2018. We carry out spectral analysis of the source in this period using both Swift/XRT and NICER observations, whereas we have performed timing analysis of NICER data only. Also, we have done broadband simultaneous spectral modelling using NICER-NuSTAR, Swift/XRT-NuSTAR and AstroSat (SXT-LAXPC) data during this period. The light curve of the source in the energy range $0.8-10$ keV, generated using both NICER and XRT observations during this period is shown in Fig.~\ref{fig:lc_all}. The NICER and XRT observations are represented using green squares and red circles, respectively. Details of the light curve are discussed in \S\ref{sec:lc-hid}. 
 Near the peak of Outburst-II, the source became extremely luminous, and XRT data suffered huge pile-up. A large area of the source region needs to be removed for pile-up correction, and it possibly severely affects the overall point spread function of the instrument. We, therefore, ignore all Swift/XRT observations between MJD 58298 to MJD 58386 from the spectral analysis.  
Similarly, the NICER instrument suffers telemetry saturation due to a very high count rate during this bright phase. Several detectors were switched off to reduce the issue of telemetry saturation.
We have considered NICER observations during this bright phase with a reduced number of active detectors and properly taken care of the reduction of effective area in each observation.

\begin{figure}
	\includegraphics[width=\columnwidth]{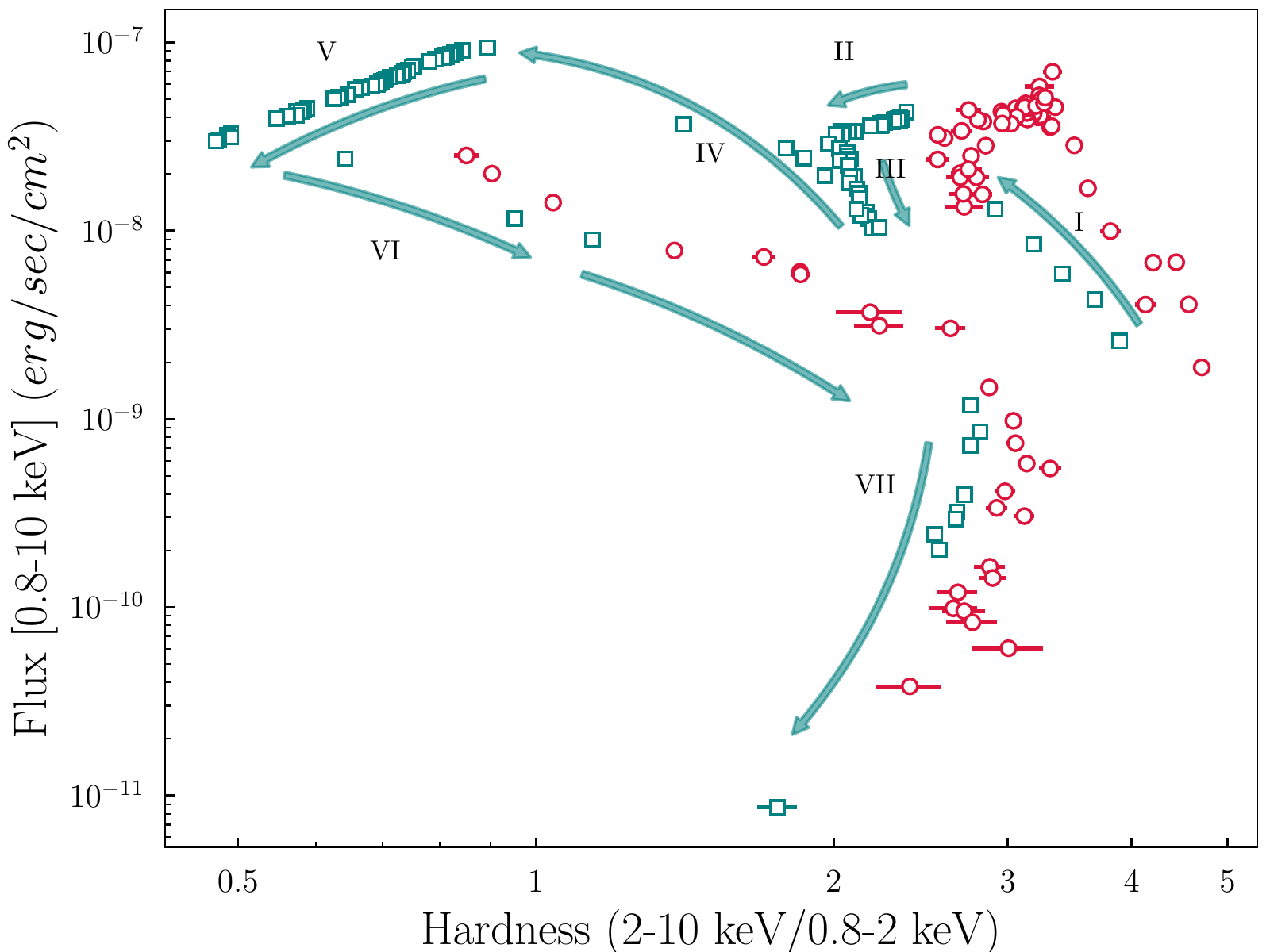}
	\caption{The Hardness-Intensity diagram of the source MAXI J1820+070 during the period MJD 58189 to MJD 58443. The hardness is calculated as the ratio of fluxes in $2-10$ keV to $0.8-2$ keV for both NICER and XRT. The red circles represent the XRT data, and green squares represent NICER data. Various phases (I-VII) of the light curve are marked. See text for details.}
	\label{fig:hid}
\end{figure}

\subsection{Spectral Analysis of Swift/XRT and NICER}
\label{sec:xrt-ni}
 We perform the spectral analysis of the source using Swift/XRT and NICER (though both operate in the similar energy range) independently to consider all the available observations in this energy range and scan the entire parameter space during both outbursts. Also, we aim for a comparative study of their spectral parameters. We use the software package \texttt{XSPEC V12.10.1f} and HEASOFT v6.26 for this analysis. The XSPEC absorption model \textit{tbabs}, which adopts the solar abundance table given by \citet{2000ApJ...542..914W}, is used to account for the interstellar (ISM) absorption. We fixed $n_H =1.5 \times 10^{21} \rm cm^{-2} $ \citep{2018ATel11423....1U} for the line of sight column density throughout the study.
The NICER and XRT observations are independently fitted in the energy range from 0.8 to 10 keV. 
The source is very bright, and we find residual in the NICER data below 0.8 keV \citep{2020MNRAS.493.5389F}. Therefore, we ignore the data below 0.8 keV for NICER and XRT to keep a common energy range for the spectral analysis.

\begin{figure}
	\centering
		\includegraphics[width=0.48\textwidth]{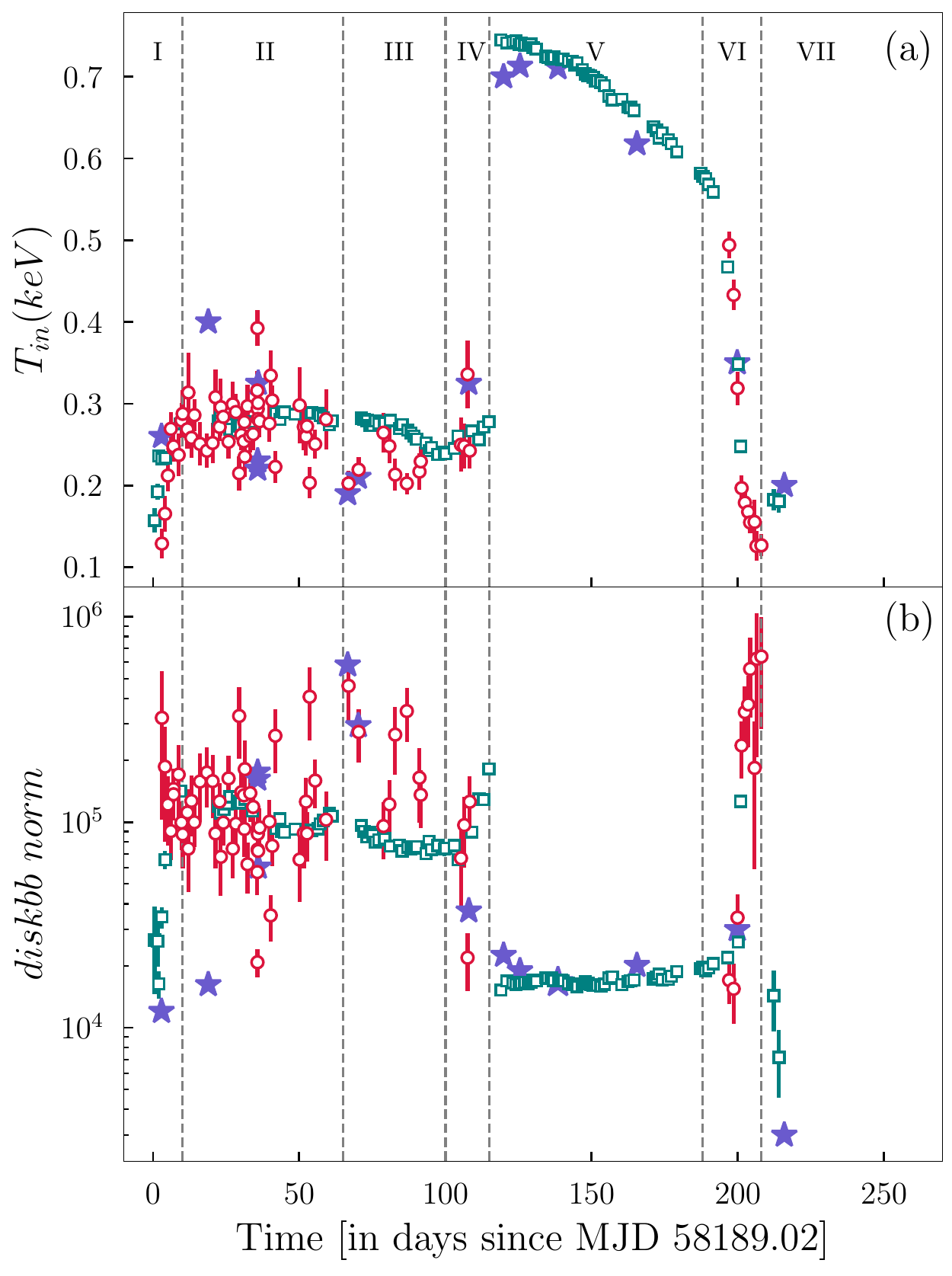} 
	\caption{Evolution of \textbf{(a)} inner accretion disc temperature ($T_{in}$) and \textbf{(b)} disk normalization for XRT and NICER data. The red circles represent the XRT data, and the green squares represent NICER data. The uncertainties are within 90$\%$ confidence range. Different evolutionary phases are marked from I to VII. The star marks represent the values estimated from the wideband spectral analysis using phenomenological models discussed in \S\ref{sec:simu-pheno}.}
	\label{fig:dbb}
\end{figure}
We consider MJD 58189.02 as day 0 of the analysis, and henceforth all results are presented with respect to this reference date. 
At the very beginning of the Outburst-I and end stages of the Outburst-II, only a \textit{powerlaw} component and the ISM absorption is enough to model (\textit{tbabs $\times$ powerlaw}) the XRT data. However, an additional multicolor blackbody (\textit{diskbb}) \citep{1984PASJ...36..741M} component is required to model (\textit{tbabs(diskbb+powerlaw)}) the XRT data in-between. NICER data show a line signature around $6-7$ keV throughout all observations of Outburst-I. Therefore, we use a Gaussian profile (\textit{gauss}) to represent the line and the overall model is \textit{tbabs(diskbb+powerlaw+gauss)} during this period. For Outburst-II, the model \textit{tbabs(diskbb+powerlaw)} is adequate for fitting the NICER data.
We use $\chi^2$ statistics to represent the goodness of the fit. The $\chi^2 /dof$  for all the XRT and NICER spectral fittings vary between $0.79-1.38$ and $0.72-1.33$ respectively.
The uncertainties in the spectral fitting parameters are calculated within 90$\%$ confidence range using the modified Levenberg-Marquardt (\textit{leven}) algorithm.
%

\subsection{Outburst Profile and HID}
\label{sec:lc-hid}
We estimate the flux in $0.8-10$ keV range for individual XRT and NICER observations during MJD 58189 to MJD 58443. 
The light curve of the source in the energy range $0.8-10$ keV, generated using both NICER and XRT observations is shown in Fig.~\ref{fig:lc_all}. The red circles represent the XRT data, and green squares represent NICER data. We notice that the overall fluxes
of both instruments are comparable. As discussed in Appendix-\ref{sec:appxa}, there may be ($10-20$\%) discrepancies in the estimated source flux due to calibration uncertainties in the spectral responses between both instruments. 
Also, a slight deviation may result due to the variation of the source itself as the XRT and NICER observations are not simultaneous. We marked different phases (I to VII) in the light curve to identify the evolution of the outburst.
Since its discovery, MAXI J1820+070 showed an increase in flux with a short rise time (Phase-I) of $\sim$ 10 days (MJD 58200) to reach the peak and then started to decay slowly (Phase-II, also referred to as plateau phase) between $\sim$ $10-60$ days (MJD 58250) followed by a much steeper decline (Phase-III) till 100 days (MJD 58290). After this phase, Outburst-II is triggered, with a rapid increase in flux (Phase-IV), and at around day 110 (MJD 58300), the X-ray flux reached the second peak ($9.38\pm 0.01 \times 10^{-8}$ erg/sec/cm$^2$), approximately 1.5 times higher than the peak value of Outburst-I in $0.8-10$ keV band. Then the source flux decays linearly (Phase-V) for the next 80 days (MJD 58380), followed by an exponential decline of flux (Phase-VI). Finally, at around day 220 (MJD 58410) the source reached the quiescence (Phase-VII).
We notice that the peak flux of Outburst-I in $2-10$ keV band is larger (a factor of 2.4 for NICER and 3 for XRT) than the flux in $0.8-2$ keV band, whereas the same for Outburst-II is lower by a factor of 1.2 (NICER). A similar trend is observed in $10-60$ keV band, represented by star symbols in Fig.~\ref{fig:lc_all}, obtained from broadband spectral studies 
(see also Swift/BAT light curve in \citet{2020MNRAS.493.5389F,2019MNRAS.490.1350B}, Insight-HXMT light curve in \citet{2021NatCo..12.1025Y,2021NatAs...5...94M}). 
Here red, green, and blue stars represent simultaneous pairs for XRT-NuSTAR, NICER-NuSTAR and AstroSat  observations, respectively.
It indicates that Outburst-I is harder in comparison with Outburst-II and we explore it in detail in \S\ref{sec:evol}.
We use these simultaneous pairs of observations for broadband spectral studies in \S\ref{sec:simu}.  
\begin{figure}
	\centering
		\includegraphics[width=0.48\textwidth]{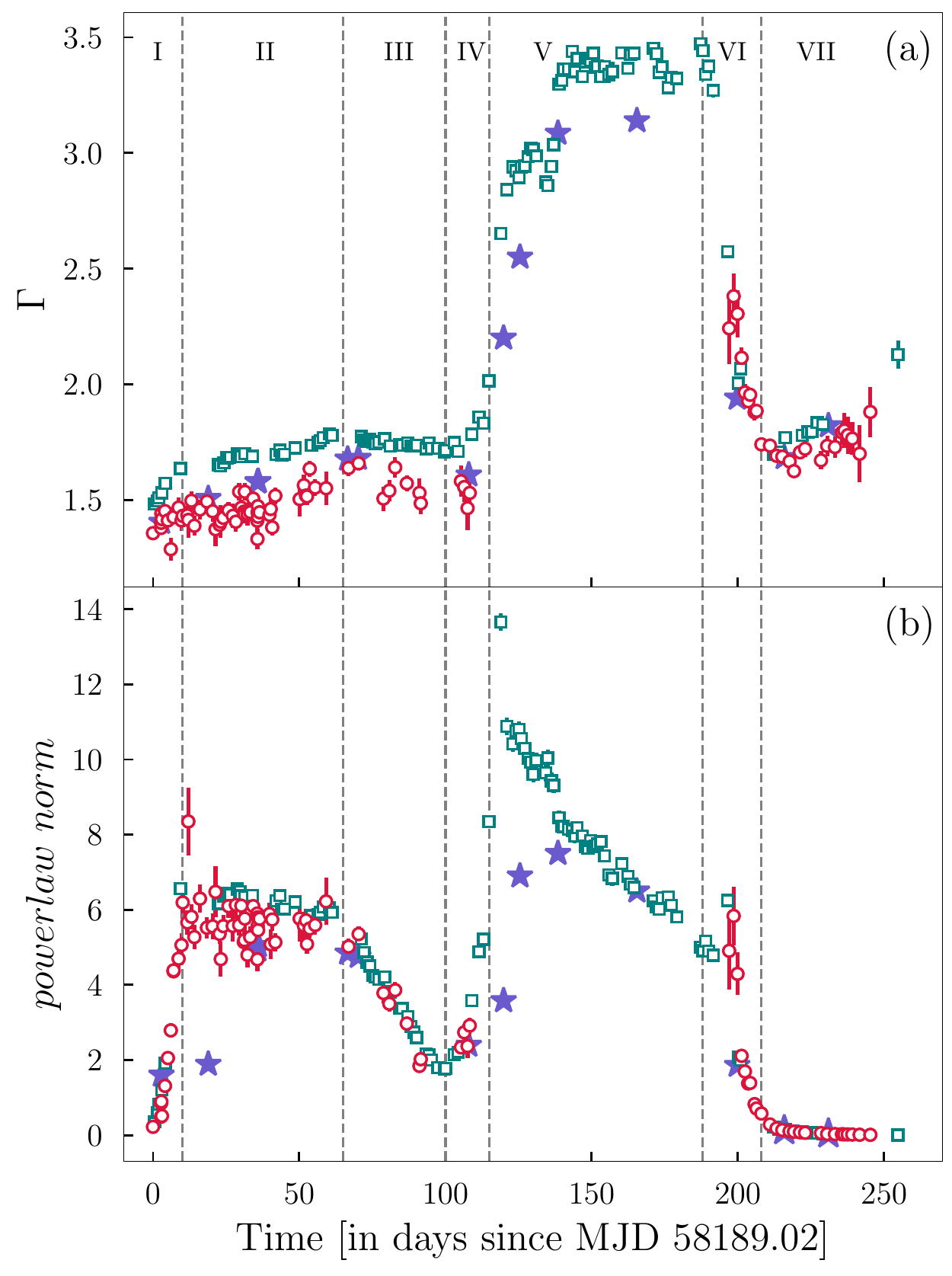}
	\caption{The time evolution of \textbf{(a)} photon index and \textbf{(b)} \textit{powerlaw} normalization for XRT and NICER data. The red circles represent the XRT data, and the green squares represent NICER data. The uncertainties are within 90$\%$ confidence range. Different evolutionary phases are marked from I to VII. The star marks represent the same estimated from the wideband spectral analysis (\S\ref{sec:simu-pheno}) using phenomenological models.}
	\label{fig:po}
\end{figure}

We also estimate the flux in $0.8-2$ keV (low) and $2-10$ keV (high) separately for all XRT and NICER observations.
We define the hardness (HR) as the ratio of flux in $2-10$ keV to that in $0.8-2$ keV. The hardness-intensity diagram for both outbursts are plotted in Fig.~\ref{fig:hid}. 
Here also, the red circles represent the XRT data points and green squares represent the NICER data points. The HID of the two instruments show an offset in HR, particularly when the source is bright during Outburst-I. In comparison with XRT, the NICER flux is $\sim 10\%$ higher in the low energy band but $\sim 10\%$ lower in the hard energy band, maintaining a comparable total flux (See Appendix-\ref{sec:appxa}). This results in a $\sim 22\%$ lower HR for NICER data than XRT.
We find that enhancing the NICER HR by this factor makes both HID comparable. We have marked various phases (I-VII) to indicate the evolution of the source in HID. 
The outburst starts with low luminosity LHS (Phase-I) with highest HR, and then evolves approximately at a constant flux level (Phase-II) with a significant change in HR. Then source flux decreases roughly at a constant HR (Phase-III) between $\sim$ $60-100$ days. 
During Outburst-I, the source remains only in the LHS (Phase-I to III in Fig.~\ref{fig:hid}) and it is a `failed' outburst. Outburst-II is triggered through Phase-IV (very short intermediate state) with an enhancement of flux, and the source reaches the HSS (Phase V).  
During the decline phase, the source returns to the LHS (Phase-VII) again via an intermediate state (Phase-VI), completing a q-shaped hysteresis cycle since the hard-to-soft transition occurs at a higher flux than soft-to-hard. It is a complete outburst since it covers all the spectral states.

\subsection{Evolution of Spectral Parameters}
\label{sec:evol}
We have modelled all the XRT and NICER data using a combination of the phenomenological models \textit{powerlaw}, \textit{diskbb} and \textit{gauss} (required only for NICER data). We aim to study the evolution of the spectral parameters during the progress of the outbursts to understand the physical scenario of the accretion process and its geometry. Also, we attempt to compare the parameters estimated using both instruments. 

We have presented the evolution of the \textit{diskbb} parameters, namely, the inner disk temperature, $T_{in}$, in Fig.~\ref{fig:dbb}a and the \textit{diskbb} normalization, \textit{diskbb norm}, in Fig.~\ref{fig:dbb}b for both XRT and NICER. XRT data points are marked by the red circles and NICER by green squares, respectively. The uncertainties are within 90$\%$ confidence range. The star marks represent the values estimated from the wideband spectral analysis using phenomenological models discussed in \S\ref{sec:simu-pheno}.  
We see that the \textit{diskbb} component is required to model all observations of Outburst-I (first $\sim 100$ days). Inner disk temperature is low at the beginning of the outburst, and it increases as the accretion of matter increases in the brief rising phase ($\sim 10$ days). Both $T_{in}$ and \textit{diskbb norm} remain almost constant (though two parameters from XRT data show some random variation) during Phase-II, although a slow decline (Phase-III) is observed towards the end of Outburst-I. It means the thermal component does not change much in Phase-II while it slowly decreases (Phase-III) in the decline part of the Outburst-I. 
The \textit{diskbb norm} is directly proportional to the inner radius of the accretion disk, $R_{in}$. The source was in the LHS during Outburst-I and the high value of \textit{diskbb norm} indicates that the disk is truncated at a large radius (see also, \citet{2021arXiv211208116Z,2021ApJ...909L...9Z}). 

Around day 100 (MJD 58290), the Outburst-II is triggered and the source becomes extremely luminous with a very sharp increase of $T_{in}$ ($\sim$ factor of 3) and order of magnitude decrease in \textit{diskbb norm}. The source transits to the HSS via a very brief intermediate state. 
The large value of $T_{in}$ signifies strong accretion activity and the low value of \textit{diskbb norm} reveals that the inner edge of the accretion disk extends close to the central object. Also, the approximately constant value of \textit{diskbb norm} during the HSS indicates that $R_{in}$ does not evolve during this period. However, $T_{in}$ gradually decreases along with the decline of the source flux. The evolution of the source in this phase is regulated by the thermal component of the radiation. As mentioned earlier, we have ignored the XRT observations during this period due to huge pile-up issues and have used NICER data for spectral modelling. 
As the source returns back to the LHS, $T_{in}$ declines to a low value, and eventually, the contribution from the disk becomes insignificant after $\sim$ 210 days (MJD 58400). The outburst ends with the disk truncated again at a large distance and \textit{diskbb norm} becomes large. 

Generally, it is proposed in the literature \citep{1980A&A....86..121S,1994MNRAS.269L..55Z,1985A&A...143..374S,1994AIPC..304..380T} that the origin of non-thermal component is due to the inverse-Comptonization of the soft thermal photons from the accretion disk by the `hot' electrons in the corona. We use a \textit{powerlaw} model to quantify the contribution of the non-thermal component.
The evolution of the \textit{powerlaw} model components are shown in Fig.~\ref{fig:po}. The XRT and NICER data are shown in red circles and green squares respectively as before. 
Fig.~\ref{fig:po}a represents the evolution of photon index ($\Gamma$) with time, whereas Fig.~\ref{fig:po}b shows the evolution of \textit{powerlaw normalization}.
Again the star marks represent the same estimated from the wideband spectral analysis using phenomenological models (See \S\ref{sec:simu-pheno}). 
In LHS, the contribution of Comptonized hard photons is larger than the soft disk photons and therefore, $\Gamma$ is low (harder spectrum) with a high value of \textit{powerlaw norm}. We notice two distinct patterns of evolution of the powerlaw component in the LHS.
Phase-II shows a gradual increase in $\Gamma$ (Fig.~\ref{fig:po}a) with a constant \textit{powerlaw norm} (Fig.~\ref{fig:po}b) such that the light curve appears as a plateau (Fig.~\ref{fig:lc_all}). However, $\Gamma$ remains constant and a gradual decline of \textit{powerlaw norm} is observed in Phase-III.
In both cases, there is a gradual decrease in the non-thermal flux. In the former case, the steepening of the spectral index due to steady thermal photon flux reduces the non-thermal flux, while in the latter case, it decreases due to the reduction of thermal photon flux without changing the nature of the non-thermal spectrum. 
Therefore, the thermal component does not change much and the non-thermal radiation regulates the dynamics in Phase-II, whereas both changes slowly in Phase-III of Outburst-I. 
We notice that $\Gamma$ (Fig.~\ref{fig:po}a) measured from NICER is higher ($\sim$ 0.24) than that of XRT during the LHS (Outburst-I) due to the difference in the instrumental response as discussed in Appendix-A. 
In Outburst-II, the spectrum is very soft ($\Gamma > 3$) as the source enters the HSS very fast and its nature remains similar during the entire HSS (Phase-V) though the non-thermal flux reduces (\textit{powerlaw norm} reduces) with time. In the declining phase, the source moves through the intermediate state ($\Gamma \sim 2$) and finally reaches the LHS (Phase-VII) with a hard spectrum.

\begin{figure}
		\includegraphics[width=0.48\textwidth]{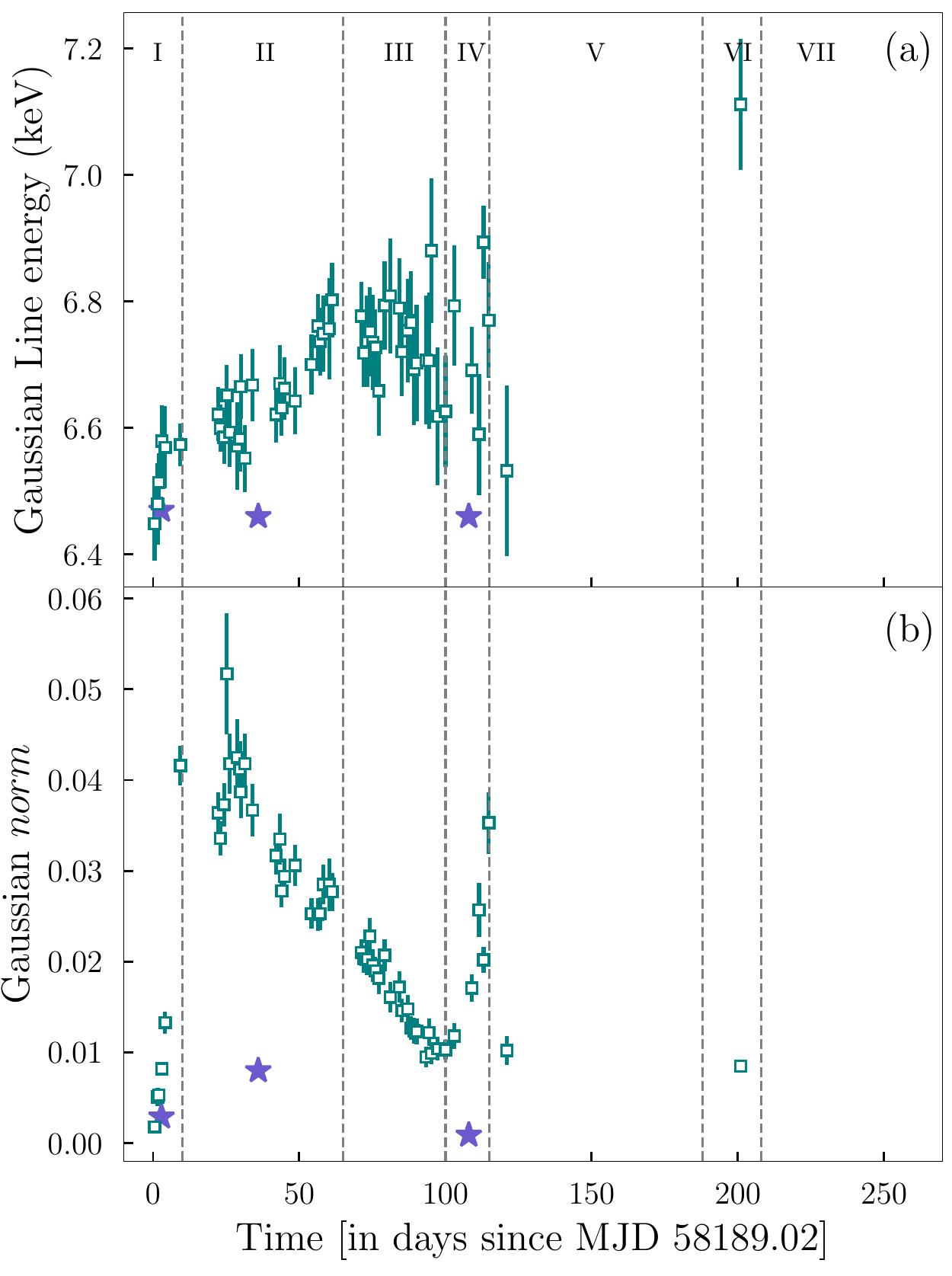}
	\caption{Time evolution of \textbf{(a)} Gaussian line energy and \textbf{(b)} Gaussian line normalization obtained from NICER data. The uncertainties are within 90$\%$ confidence range. Different evolutionary phases are marked from I to VII. The star marks represent the parameters estimated from the wideband spectral analysis (\S\ref{sec:simu-pheno}) using phenomenological models.}
	\label{fig:gauss}
\end{figure} 
The spectral modelling of NICER data show an excess residual in $6-7$ keV range which is generally assumed to be Fe $K_{\alpha}$ feature 
produced due to the reflection of hard photons onto the cold accretion disk. We use a Gaussian profile wherever the feature is visible.  The evolution of Fe $K_{\alpha}$ line energy (Fig.~\ref{fig:gauss}a) and the \textit{Gaussian normalization} (Fig.~\ref{fig:gauss}b) estimated using NICER data are represented by green square. The star marks show the same from the broadband spectral modelling. The Gaussian line width for NICER is frozen at 0.6 keV for some observations, which unless giving very high values.  
At the beginning of the outburst, the source is in LHS and more hard \textit{powerlaw} photons from the corona illuminate the accretion disk surface. We notice an increasing line contribution during the rising phase. Then there is a gradual decrease of \textit{powerlaw} photons which results in a decrease in the amount of reflected radiation. This trend can be clearly visible in Fig.~\ref{fig:gauss}b. There are 
other factors, like, the slope of the hard photon spectrum from the corona, the ionization of the disk surface, etc. \citep{2019MNRAS.490.1350B,2019Natur.565..198K,2021MNRAS.507.2744A} are also responsible for the observed changes in the iron line profile.
The Fe $K_{\alpha}$ line emission energy is 6.4 keV but it can vary slightly due to the relativistic effects.
Fig.~\ref{fig:gauss}a shows the systematic increase of the Gauss line energy and it lies in the range $6.53-6.89$ keV. In Outburst-II, the source immediately enters the HSS, the disk photons dominate and further no more line features visible in the spectra. Although the NICER observation around $\sim$ 200 day shows a reflection feature at $\sim$ 7.1 keV.
 
\begin{table*}
	\centering
	\caption{Simultaneous broadband (XRT-NuSTAR, NICER-NuSTAR and AstroSat) observations of the source MAXI J1820+070}
	\label{tab:simult-obs}
	\begin{tabular}{|l|c|c|c|c|r|}  
		\hline
		\multirow{2}{*}{Epoch} & \multicolumn{3}{c}{Obs. ID (MJD)} & \\
		\cline{2-5} 
		  & NuSTAR & XRT & NICER & AstroSat\\
		\hline
		1 & 90401309002 (58191.94) & & 1200120103 (58190.99) & \\ 
		2 & & & & 9000001994 (58208) \\
		3 & 90401309013 (58224.95) & 00010627042 (58224.87) & & \\ 
		4 & 90401309013 (58224.95) & 00010627043 (58224.95) & &\\ 
		5 & 90401309014 (58225.28) & & 1200120131 (58225.03) & \\ 
		6 & 90401309019 (58255.60) & 00088657004 (58255.90) & & \\
		7 & 90401324002 (58259.24) & 00010627062 (58259.35) & & \\
		8 & 90401309021 (58297.17) & & 1200120189 (58297.19) &\\ 
		9 & & & & 9000002216 (58309) \\
		10 & 90401309025 (58314.75) & & 1200120207 (58314.19) & \\ 
		11 & 90401309027 (58327.05)	& &	1200120220 (58327.89) & \\ 
		12 & & & & 9000002324 (58355) \\
		13 & 90401309033 (58388.91) & 00088657010 (58388.92) & & \\ 
		14 & 90401309037 (58404.95) & & 1200120277 (58405.20)& \\ 
		15 & 90401309039 (58420.05) & & 1200120290 (58420.05) & \\ 
		\hline
	\end{tabular}
\end{table*}
\subsection{Simultaneous Wideband Spectral Modelling}
\label{sec:simu}
We have performed the simultaneous wideband spectral analysis of four different instruments, Swift/XRT, NICER, NuSTAR, and AstroSat to understand the spectral characteristics over a broad range.
We consider all simultaneous or quasi-simultaneous (with a time difference not exceeding 1 hour) observations (XRT-NuSTAR, NICER-NuSTAR and AstroSat SXT-LAXPC) which are listed in Table~\ref{tab:simult-obs}. We used XRT or NICER data for the low energy band ($0.8-10$ keV), and NuSTAR data ($4-75$ keV) for the high energy regime. For AstroSat, we use SXT ($0.7-8$ keV) and LAXPC ($3-60$ keV) for the broadband simultaneous spectral modelling. 
The SXT data below 1.3 keV for Obs. ID 9000001994 (Epoch 2 in Table~\ref{tab:simult-obs}) show a high systematic residual, which is also reported by \citet{2020MNRAS.498.5873C}. 
Therefore, we have used SXT data between 1.3 - 7.0 keV for this observation.
Temporal filter is applied to the simultaneous NuSTAR data according to the exposure time of the corresponding XRT/NICER observation to ensure they are strictly simultaneous. The spectra from both focal plane modules (FPMA and FPMB) of NuSTAR data are fitted together. A normalization constant is used to account for the calibration difference, if any, between FPMA and FPMB.
 
We consider 15 epochs (5 pairs of XRT-NuSTAR, 3 pairs of AstroSat and 7 pairs of NICER-NuSTAR) of simultaneous data (Table~\ref{tab:simult-obs}) and model them using combination of phenomenological models for a comparison of parameters obtained from the spectral modelling of individual NICER/XRT data. 
Also, we model the broadband data using reflection model to estimate the accretion disk characteristics, BH spin, inclination angle etc. The $n_H$ value is fixed at 1.5$\times 10^{21} \rm cm^{-2}$, same as the XRT/NICER spectral modelling (\S\ref{sec:xrt-ni}). 
AstroSat/LAXPC data show an instrumental feature $\sim 30$ keV due to Xe K fluorescence X-rays since the detector is filled with a mixture of xenon and methane \citep{2017ApJS..231...10A,2021JApA...42...32A,2019MNRAS.487..928S}.  
Additional absorption \textit{edges} are used in order to address this instrumental feature of LAXPC.
 
\definecolor{Gray}{gray}{0.9}
\definecolor{silver}{rgb}{0.75, 0.75, 0.75}
\begin{table*}
	\centering
	\caption{Spectral parameters of simultaneous broadband X-ray data (XRT-NuSTAR, NICER-NuSTAR and AstroSat) fitted with phenomenological models. The XRT-NuSTAR pairs and AstroSat observations are highlighted with dark-grey and pale-gray colors, respectively; others are NICER-NuSTAR pairs. The uncertainties are within 90$\%$ confidence range.}
	\label{tab:simult-pheno}
	\begin{tabular}{|c|c|c|c|c|c|c|c|c|c|c|}  
		\hline
	    \multirow{3}{*}{Epoch} & \multicolumn{8}{c}{Model} &  & \multirow{3}{*}{$\chi^2_{red}$} \\
	    \cline{2-10} 
	     & \multicolumn{2}{c|}{\textit{diskbb}} & \multicolumn{3}{c|}{\textit{gauss}} & \multicolumn{2}{c|}{\textit{smedge}} & \multicolumn{2}{c|}{\textit{powerlaw}} & \\
	     \cline{2-10}
	     & $T_{in}$ & \textit{norm} &  \textit{line E} & \textit{sigma} & \textit{norm} & \textit{edge E} & \textit{width} & $\Gamma $ & \textit{norm} &  \\
	     & (\textit{keV}) & ($ \times 10^4$) & (\textit{keV}) & (\textit{keV}) & \textit{(photons/cm$^2$/s)}& (\textit{keV}) & (\textit{keV}) & &\textit{(photons/keV/cm$^2$/s)} &\\
	    
		\hline
		&&&&&&&&&& \\
		1  & $0.26_{-0.01}^{+0.01}$ & $1.2_{-0.3}^{+0.3}$ & $6.47_{-0.04}^{+0.04}$ & $0.25_{-0.05}^{+0.05}$ & $0.0029_{-0.0005}^{+0.0006}$ & $6.57_{-0.09}^{+0.08}$ & $8.9_{-1.7}^{+2.4}$ & $1.403_{-0.002}^{+0.002}$ & $1.6_{-0.3}^{+0.4}$ & 1.17 \\ 
		&&&&&&&&&& \\
		 \rowcolor{Gray}
		2 & $0.391_{-0.002}^{+0.003}$ & $1.62_{-0.13}^{+0.09}$ & ---& --- &--- & $8.1_{-0.3}^{+0.3}$ & $3.8_{-0.9}^{+0.4}$ & $1.506_{-0.001}^{+0.009}$ & $4.37_{-0.06}^{+0.06}$ & 1.43 \\
		&&&&&&&&&& \\
		\rowcolor{silver}
		3 & $0.23_{-0.01}^{+0.01}$ 	&  $17.5_{-4.8}^{+6.9}$ 	& ---		 				& ---						& ---												& $7.24_{-0.09}^{+0.09}$	& $3.7_{-0.6}^{+0.8}$	& $1.578_{-0.003}^{+0.003}$	& $4.93_{-0.03}^{+0.03}$		& 1.33 \\ 
		&&&&&&&&&& \\
		\rowcolor{silver}
		4 & $0.22_{-0.02}^{+0.02}$ 	& $16.2_{-5.3}^{+8.3}$ 		& ---						& ---						& ---												& $7.2_{-0.1}^{+0.1}$		& $3.7_{-0.6}^{+0.8}$	& $1.578_{-0.003}^{+0.003}$	& $4.94_{-0.03}^{+0.03}$		& 1.28 \\ 
		&&&&&&&&&& \\
		5 & $0.325_{-0.003}^{+0.003}$ 	& $6.04_{-0.2}^{+0.2}$ 		& $6.46_{-0.03}^{+0.03}$	& $0.28_{-0.03}^{+0.03}$	& $0.0083_{-0.0008}^{+0.0008}$	& $7.33_{-0.06}^{+0.06}$	& $3.2_{-0.3}^{+0.4}$	& $1.580_{-0.001}^{+0.001}$	& $5.05_{-0.01}^{+0.01}$		& 1.89 \\ 
		&&&&&&&&&& \\
		\rowcolor{silver}
		6 & $0.19_{-0.01}^{+0.01}$ & $58.2_{-18.6}^{+28.9}$ & --- & --- & --- & $7.2_{-0.1}^{+0.1}$ & $4.4_{-1.1}^{+1.6}$ & $1.677_{-0.004}^{+0.004}$ & $4.85_{-0.04}^{+0.04}$ & 1.06 \\		
		&&&&&&&&&& \\
		\rowcolor{silver}
		7 & $0.21_{-0.01}^{+0.01}$ & $29.5_{-7.6}^{+10.7}$ & --- &--- & --- & $7.2_{-0.2}^{+0.2}$ & $4.8_{-1.5}^{+2.4}$ & $1.68_{-0.01}^{+0.01}$ & $4.76_{-0.04}^{+0.04}$ & 1.10 \\
		
		&&&&&&&&&& \\

		8 & $0.325_{-0.002}^{+0.002}$ 	& $3.7_{-0.1}^{+0.1}$		& $6.46_{-0.04}^{+0.04}$	& $0.12_{-0.05}^{+0.05}$	& $0.0009_{-0.0002}^{+0.0002}$	& $7.33_{-0.04}^{+0.04}$	& $5.1_{-0.4}^{+0.5}$	& $1.608_{-0.001}^{+0.001}$	& $2.395_{-0.006}^{+0.006}$		& 1.55 \\ 
		&&&&&&&&&& \\
		\rowcolor{Gray}
		9 & $0.7_{-1.1}^{+0.6}$ & $2.25_{-0.02}^{+0.03}$ &--- &--- &--- & $7.58_{-0.07}^{+0.07}$ & $16.5_{-1.1}^{+0.6}$ & $2.20_{-0.01}^{+0.01}$ & $3.58_{-0.08}^{+0.09}$ & 1.7 \\
		&&&&&&&&&& \\
		10 &  $0.713_{-0.001}^{+0.001}$	& $1.89_{-0.02}^{+0.02}$	& ---						& ---						& ---												& $6.87_{-0.05}^{+0.06}$	& $14.3_{-3.1}^{+3.2}$	& $2.55_{-0.01}^{+0.02}$	& $6.9_{-0.2}^{+0.2}$			& 1.42 \\ 
		&&&&&&&&&& \\
		11 & $0.711_{-0.001}^{+0.001}$& $1.66_{-0.02}^{+0.02}$	& ---						& ---						& ---												& $6.76_{-0.08}^{+0.08}$	& $13.9_{-0.4}^{+0.9}$	& $3.086_{-0.017}^{+0.006}$	& $7.5_{-0.2}^{+0.2}$			& 1.09 \\ 
		&&&&&&&&&& \\
		\rowcolor{Gray}
		12 & $0.618_{-0.002}^{+0.002}$ & $2.01_{-0.04}^{+0.04}$ & --- & --- & --- & $8.9_{-0.3}^{+0.4}$ & $8.84^{\dag}$ & $3.14_{-0.02}^{+0.02}$ & $6.5_{-0.2}^{+0.2}$ & 1.80  \\
		&&&&&&&&&& \\
		\rowcolor{silver}
		13 & $0.35_{-0.01}^{+0.01}$ 	& $3.0_{-0.3}^{+0.4}$ 		& ---						& ---						& ---												& $7.8_{-0.2}^{+0.2}$		& $13.87^{\dag}$				& $1.94_{-0.02}^{+0.01}$	& $1.86_{-0.05}^{+0.05}$		&  1.03 \\ 
		&&&&&&&&&& \\
		14 & $0.20_{-0.02}^{+0.02}$ 	& $0.3_{-0.1}^{+0.2}$		& ---						& ---						& ---												& ---						&---					& $1.683_{-0.009}^{+0.009}$	& $0.124_{-0.003}^{+0.003}$		& 0.97 \\ 
		&&&&&&&&&& \\
		15 &  --- 						& --- 						& ---						& ---						& ---												& ---						&---					& $1.821_{-0.006}^{+0.006}$	& $0.0383_{-0.0005}^{+0.0006}$& 1.00 \\ 
		&&&&&&&&&& \\
		\hline
	\end{tabular}			
	\begin{tablenotes}
			\item[b] \dag Parameter uncertainty can't be estimated.
		\end{tablenotes}  
\end{table*}
\begin{figure*}   
	\centering
		\includegraphics[width=0.30\textwidth,angle =-90]{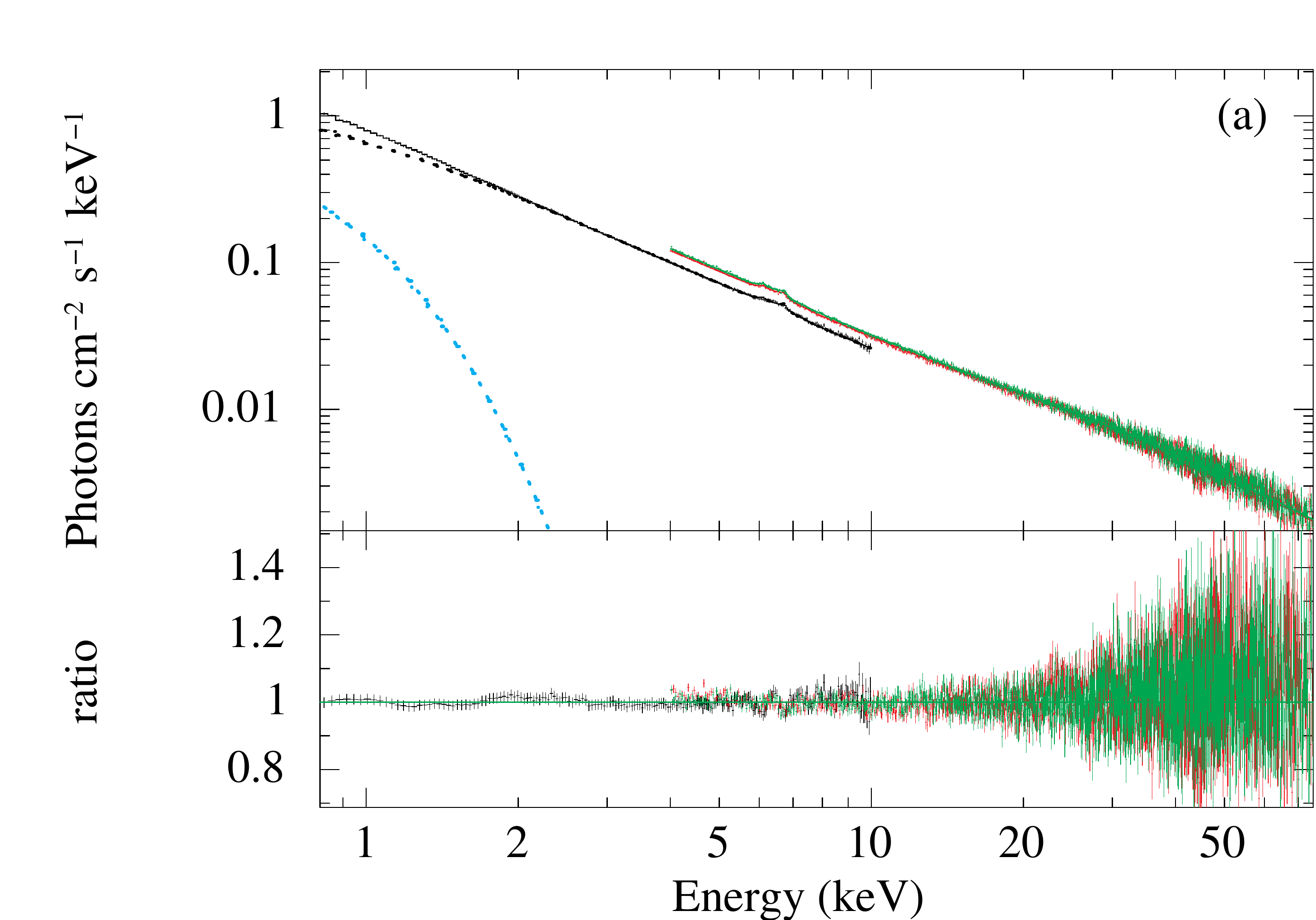}
		\includegraphics[width=0.30\textwidth,angle =-90]{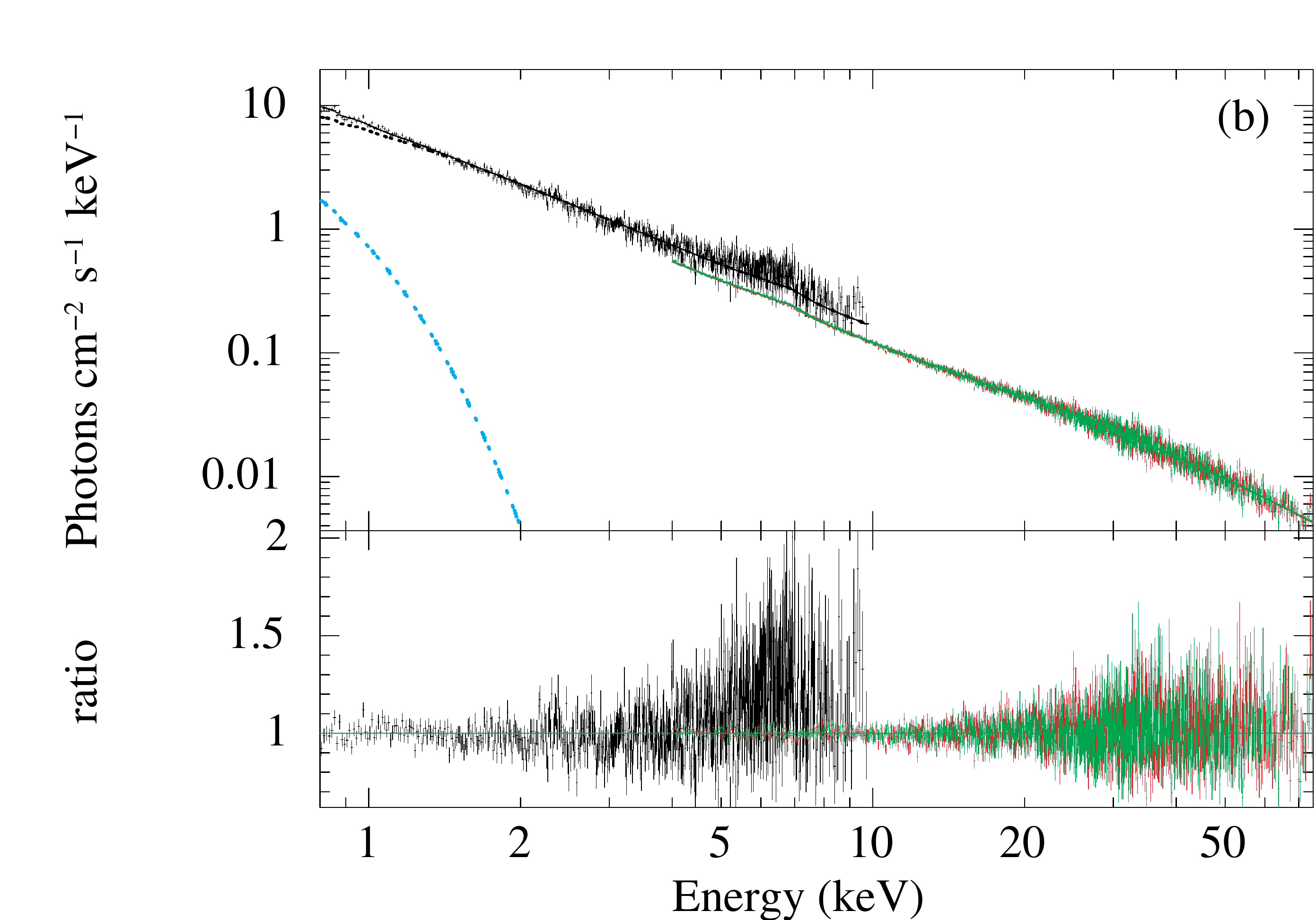}
	\caption{\textbf{(a)} Simultaneous NICER-NuSTAR observation at MJD 58191 (Epoch 1 in Table~\ref{tab:simult-obs}) fitted using the model \textit{tbabs(diskbb+relxilllpCp)}. Black color represents NICER data and red and green for NuSTAR FPMA and FPMB respectively. \textbf{(b)} Quasi-Simultaneous XRT-NuSTAR observation at MJD 58224 (Epoch 3 in Table~\ref{tab:simult-obs}) fitted using the model \textit{tbabs(diskbb+relxilllpCp)}. Black color represents XRT data and red and green for NuSTAR FPMA and FPMB, respectively. The \textit{diskbb} and reflection components are shown in light blue and black dotted lines, respectively.}
	\label{fig:sim}
\end{figure*}   
 
\subsubsection{Phenomenological Spectral Modelling}
\label{sec:simu-pheno}
We have attempted to model the XRT-NuSTAR, NICER-NuSTAR and AstroSat observations using a combination of the phenomenological models 
\textit{diskbb} and \textit{nthcomp} with a motivation to estimate the temperature of the corona and exponential cut-off, if any. But we
could neither constrain the electron temperature nor the cut-off energy. Therefore, we continue to model the broadband data with \textit{diskbb} and \textit{powerlaw} to maintain a consistency with the spectral modelling of individual NICER/XRT data. 
Additionally, we have used components like \textit{gauss} and \textit{smedge} if they are required. The estimated parameters are presented in Table~\ref{tab:simult-pheno} and marked with star symbol in Fig.~\ref{fig:dbb},\ref{fig:po},\ref{fig:gauss}. {The \textit{diskbb} parameters are consistent with NICER/XRT, though Epoch 2 (AstroSat) records a higher $T_{in}$ and lower \textit{diskbb norm} than NICER/XRT}. Epoch $1-8$ and Epoch $14-15$ in Table~\ref{tab:simult-pheno} belong to LHS and the estimated \textit{powerlaw} index is consistent with the same from XRT data alone (Fig. \ref{fig:po}a). Epoch $9-12$ corresponds to HSS and the estimated $\Gamma$ ($\lesssim 3$) is significantly harder than that estimated from the NICER data (Fig. \ref{fig:po}a). During Epoch 13, the source was possibly transiting through an intermediate state with  $\Gamma=1.94_{-0.02}^{+0.01}$.
Additional \textit{Gausian} component is required for three broadband (Epoch 1, 5 \& 8 in Table \ref{tab:simult-pheno}) observations and the line parameters differ significantly from that of NICER data (Fig.~\ref{fig:gauss}). This is because NuSTAR has a broad coverage to capture the reflection features and has a better energy resolution in this range. Inclusion of NuSTAR data results in constraining the \textit{gauss} line energy and width (Table \ref{tab:simult-pheno}) better than NICER data alone.  
We also notice the presence of absorption edge ($6.75-9.3$ keV) in all the simultaneous data except Epochs 14 and 15 (towards the quiescent state). It possibly indicates that significant amount of hard X-ray irradiation is present during the outbursts. 

\subsubsection{Spectral Modelling for Reflection Features}
\label{sec:simu-refl}
The hard X-ray photons from the corona can irradiate the accretion disk surface, getting reprocessed by absorption and down scattering, producing a reflection signature in the observed data. 
Therefore, X-ray reflection spectroscopy is a powerful tool for studying the accretion disk characteristics such as the degree of ionization and the elemental abundances. 
The reflection lines suffer relativistic broadening and its shape depends on spin of the BH, the inclination of the accretion disk, geometry of the corona, etc. \citep{2013MNRAS.430.1694D}.
Therefore, the study of the relativistic reflected spectrum is a direct probe towards the physics around the strong gravity region
and we can determine the spin of the BH and the inclination of the system, in addition to the accretion disk characteristics. 
Generally, the reflection models assume a primary spectrum, either a broken powerlaw emissivity (in case of extended corona) or an isotropic point-like source at a given height (lamp-post corona) on the rotation axis of the BH. 
There are studies related to reflection features in the spectra of MAXI J1820+070 on a selected number of observations using 
various flavors of reflection models \citep{2019MNRAS.490.1350B,2020MNRAS.498.5873C,2021MNRAS.500.3976B,2021A&A...656A..63M,2021NatCo..12.1025Y,2022MNRAS.tmp..101K}.

In our study, we have performed broadband spectral modelling using a few different flavors of the relativistic reflection model RELXILL\footnote{\url{http://www.sternwarte.uni-erlangen.de/~dauser/research/relxill/index.html}} such as \textit{relxilllpCp}, \textit{relxilllp} \citep{2014ApJ...782...76G}. This model uses the transfer function method  \citep{1975ApJ...202..788C} to calculate the photon trajectory from the point of emission in the accretion disk to the observer (a \textit{ray-tracing code}). The model takes care of the proper emission angle at each point on the disk and calculates the reflected flux, instead of following an angle-averaged method. 
The \textit{relxilllpCp} model assumes a lamp-post geometry for the corona and \textit{nthcomp} continuum \citep{2014MNRAS.444L.100D} for the incident primary spectrum, whereas in \textit{relxilllp} model a cut-off \textit{powerlaw} continuum is considered as the primary spectrum with the same coronal geometry.

\begin{figure}
	\includegraphics[width=0.31\textwidth,angle =-90]{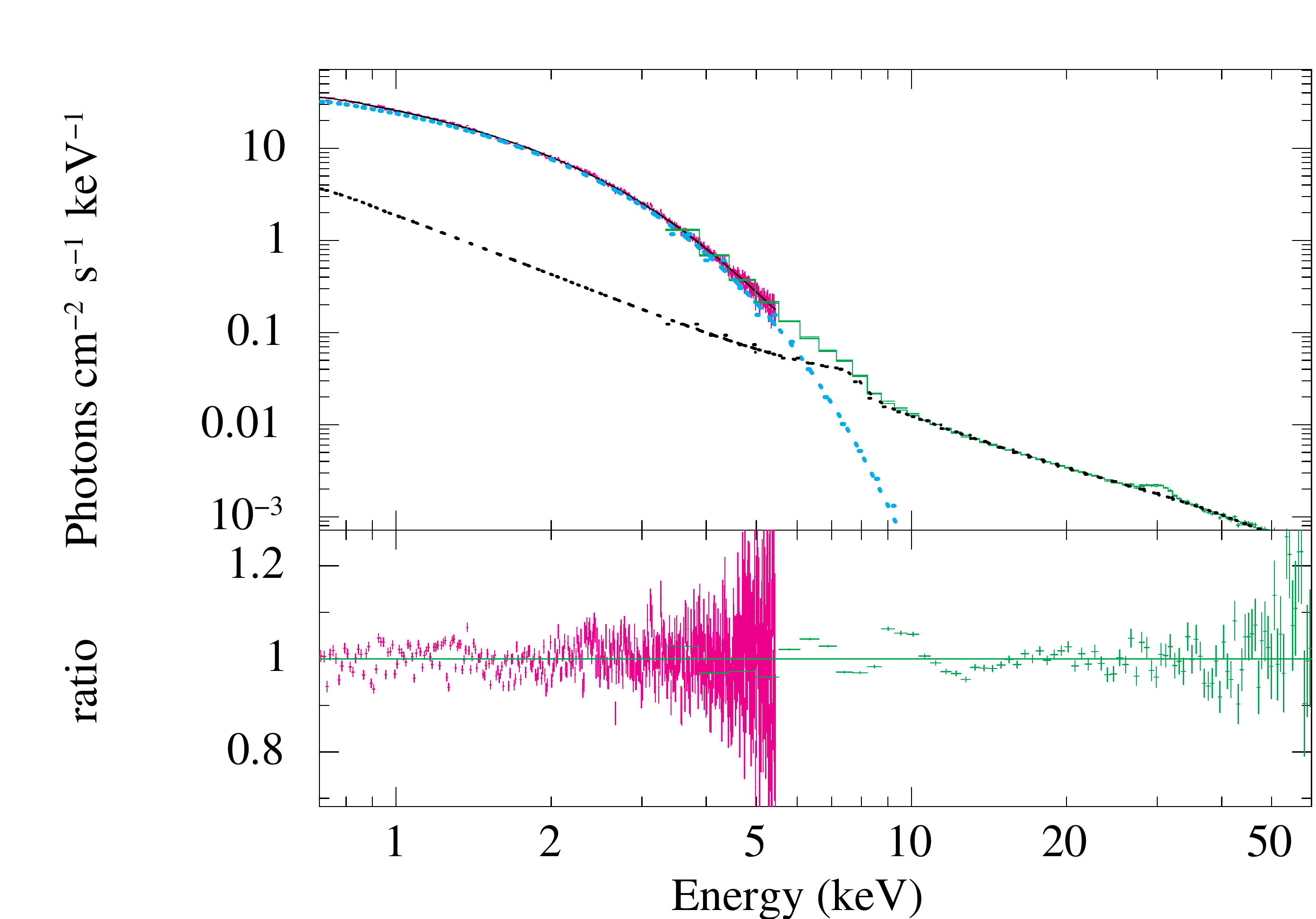}
	\caption{Broadband energy spectrum of AstroSat observation (Epoch 9) is best fitted with the model \textit{tbabs(diskbb+relxilllp)}. The SXT and LAXPC data are plotted in magenta and green color, respectively. The \textit{diskbb} and reflection components are shown in light blue and black dotted lines respectively.}
	\label{fig:astro-refl}
\end{figure}

\begin{table*}
	\centering
	\caption{Spectral parameters of simultaneous broadband XRT-NuSTAR, NICER-NuSTAR and AstroSat observations fitted using RELXILL group of models. The XRT-NuSTAR pairs and AstroSat observations are highlighted with dark-grey and pale-gray colors respectively, others are NICER-NuSTAR pairs. The uncertainties are within 90$\%$ confidence range.}
	\label{tab:simult-refl}
	\begin{tabular}{|c|c|c|c|c|c|c|c|c|c|c|c|}  
		\hline
	\multirow{3}{*}{Epoch} & \multicolumn{10}{c|}{Model} & \multirow{3}{*}{$\chi^2_{red}$} \\
	\cline{2-11} 
	& \multicolumn{2}{c|}{\textit{diskbb}}  & \multicolumn{8}{c|}{\textit{relxilllpCp}} & \\
	\cline{2-11} 
    & $T_{in}$ & \textit{norm} & \textit{h} & $\theta$  & $\Gamma$ & $log$ $\xi $ & $A_{Fe}$ & $R_f$ & $kT_e $ & \textit{norm} & \\
    & \textit{(keV)} & ($ \times 10^4$) & $(GM/c^2)$ & (\textit{deg}) & & (\textit{erg cm s$^{-1}$}) & ($A_{Fe,\odot}$) & & \textit{(keV)} & ($\times 10^{-2}$)& \\ 
		\hline
		&&&&&&&&&&& \\
		1 & $0.237_{-0.002}^{+0.007}$	& $1.59_{-0.2}^{+0.2}$		& $202.25_{-64.2}^{+159.01}$	&$63.0^{\dag \dag}$		&$1.553_{-0.001}^{+0.003}$	&$2.01_{-0.06}^{+0.08}$	&$2.4_{-0.1}^{+0.1}$	&$0.39_{-0.01}^{+0.01}$	&$100.0^{\dag \dag}$	&$5.9_{-0.1}^{+0.1}$	& 1.09 \\ 
		&&&&&&&&&&& \\
		\rowcolor{Gray}
		2 & $0.33_{-0.02}^{+0.02}$ & $2.3_{-0.5}^{+0.6}$ & $100.0^{\dag \dag}$ & $63.0^{\dag \dag}$ & $1.54_{-0.01}^{+0.01}$ & 
$3.70_{-0.04}^{+0.07}$ & $4.4_{-0.8}^{+2.5}$ & $0.7_{-0.3}^{+0.2}$ & $100.0^{\dag \dag}$ & $16.5_{-1.7}^{+2.8}$ & 1.46 \\
		&&&&&&&&&&& \\
		\rowcolor{silver}
		3 & $0.17_{-0.02}^{+0.02}$		& $44.6_{-22.5}^{+51.8}$	& $29.46_{-9.55}^{+10.36}$		&$56.9_{-7.0}^{+9.6}$	&$1.72_{-0.01}^{+0.01}$		&$1.69^\dag$			&$2.4_{-0.3}^{+0.3}$	&$0.6_{-0.1}^{+0.2}$	&$116.67^\dag$	&$18.2_{-0.9}^{+1.5}$	&1.20 \\ 
		&&&&&&&&&&& \\
		\rowcolor{silver}
		4 & $0.17_{-0.02}^{+0.02}$		& $47.1_{-27.0}^{+53.6}$	& $30.44_{-4.84}^{+6.88}$		&$59.8_{-3.7}^{+4.3}$	&$1.73_{-0.01}^{+0.01}$		&$1.70_{-0.09}^{+0.05}$	&$2.3_{-0.1}^{+0.4}$	&$0.61_{-0.09}^{+0.14}$	&$100.0^{\dag \dag}$	&$17.6_{-0.7}^{+0.8}$	&1.14 \\ 
		&&&&&&&&&&& \\
		5 & $0.296_{-0.003}^{+0.002}$	& $7.02_{-0.4}^{+0.5}$		& $30^\dag$						&$63.0^{\dag \dag}$		&$1.729_{-0.003}^{+0.002}$	&$1.70_{-0.02}^{+0.01}$	&$2.14_{-0.06}^{+0.09}$	&$0.67_{-0.01}^{+0.01}$	&$100.0^{\dag \dag}$	&$18.0_ {-0.1}^{+0.1}$	&1.13 \\ 
		&&&&&&&&&&& \\
		\rowcolor{silver}
		6 & $0.16_{-0.01}^{+0.01}$ & $105.9_{-53.7}^{+100.4}$ & $23.1_{-3.3}^{+4.5}$ & $63.0^{\dag \dag}$ & $1.808_{-0.004}^{+0.004}$ & $1.70_{-0.03}^{+0.02}$ & $2.7_{-0.2}^{+0.4}$ & $0.65_{-0.03}^{+0.02}$ & $100.0^{\dag \dag}$ & $13.7_{-0.3}^{+0.2}$ & 1.08 \\
		&&&&&&&&&&& \\
		\rowcolor{silver}
		7 & $0.17_{-0.01}^{+0.01}$ & $58.4_{-21.6}^{+22.9}$ & $23.1_{-4.6}^{+5.5}$ & $63.0^{\dag \dag}$ & $1.82_{-0.01}^{+0.01}$ & $1.70_{-0.04}^{+0.02}$ & $2.7_{-0.2}^{+0.5}$ & $0.67_{-0.03}^{+0.06}$ & $100.0^{\dag \dag}$ & $13.2_{-0.2}^{+0.3}$ & 1.11 \\		
		&&&&&&&&&&& \\
		8 & $0.306_{-0.001}^{+0.003}$	& $4.0_{-0.1}^{+0.2}$		& $20.3_{-1.2}^{+0.6}$			&$63.0_{-0.6}^{+0.4}$	&$1.741_{-0.001}^{+0.001}$	&$1.699_{-0.003}^{+0.003}$&$2.81_{-0.07}^{+0.05}$&$0.59_{-0.02}^{+0.03}$&$100.0^{\dag \dag}$	&$8.57_{-0.03}^{+0.05}$	&1.52 \\ 
		&&&&&&&&&&& \\
		\rowcolor{Gray}
		9$^*$ & $0.710_{-0.002}^{+0.002}$ & $1.91_{-0.03}^{+0.04}$ & $3.0_{-0.5}^{+0.5}$ & $63.0^{\dag \dag}$ & $2.32_{-0.03}^{+0.02} $ & $3.09_{-0.03}^{+0.04}$ & $10.0^{\ddag}$ & $3.6_{-0.2}^{+0.3} $ & --- & $14.8_{-4.4}^{+8.8} $ & 1.58  \\
		&&&&&&&&&&& \\
		10$^*$ & $0.7398_{-0.001}^{+0.001}$	& $1.67_{-0.02}^{+0.02}$	& $2.0^{\ddag}$					&$63.0^{\dag \dag}$		&$2.57_{-0.01}^{+0.01}$		&$3.31_{-0.01}^{+0.02}$	&$7.9_{-1.5}^{+1.1}$	&$3.10_{-0.03}^{+0.27}$	&---			&$125.3_{-1.1}^{+9.0}$	&1.19 \\ 
		&&&&&&&&&&& \\
		11$^*$ & $0.7198_{-0.001}^{+0.001}$	& $1.67_{-0.003}^{+0.003}$	& $2.23_{-0.02}^{+0.04}$		&$59.9_{-0.4}^{+0.3}$	&$3.273_{-0.050}^{+0.003}$	&$2.70_{-0.02}^{+0.01}$	&$10.0^{\ddag}$			&$3.8_{-0.1}^{+0.1}$	&---			&$424.5_{-4.6}^{+126}$	&1.28 \\ 
		&&&&&&&&&&& \\
		\rowcolor{Gray}
		12$^*$ & $0.691_{-0.002}^{+0.002}$ & $0.55_{-0.01}^{+0.01}$ & $2.69_{-0.45}^{+0.45}$ & $63.0^{\dag \dag}$ &  $3.12_{-0.01}^{+0.01}$  & $4.3_{-0.06}^{+0.06}$ & $10.0^{\ddag}$ & $3.59_{-0.37}^{+0.37}$ & --- & $15.8_{-1.0}^{+1.0}$ & 1.37
 \\
		&&&&&&&&&&& \\
		\rowcolor{silver}
		13 & $0.35_{-0.01}^{+0.01}$	& $2.38_{-0.4}^{+0.4}$		& $11.25_{-2.26}^{+3.65}$		&$63.0^{\dag \dag}$		&$2.07_{-0.02}^{+0.02}$		&$1.7_{-0.1}^{+0.2}$	&$4.7_{-0.7}^{+0.4}$	&$0.8_{-0.1}^{+0.1}$	&$100.0^{\dag \dag}$	&$3.9_{-0.2}^{+0.5}$	& 1.10\\ 
		&&&&&&&&&&& \\
		14 & ---						& ---						& $8.6^\dag$					&$63.0^{\dag \dag}$		&$1.73_{-0.01}^{+0.01}$		&$2.8_{-0.1}^{+0.1}$	&$1.4_{-0.5}^{+0.7}$	&$0.38_{-0.04}^{+0.05}$	&$54.34^\dag$	&$0.5_{-0.1}^{+0.2}$	&1.0 \\ 
		&&&&&&&&&&& \\
		15 & ---						& ---						& $3.03^{\dag}$				&$63.0^{\dag \dag}$		&$1.84_{-0.01}^{+0.01}$		&$1.7_{-0.6}^{+0.2}$	&$5.0_{-1.2}^{+2.4}$	&$0.38_{-0.07}^{+0.07}$	&$100.0^{\dag \dag}$	&$0.38_{-0.004}^{+0.003}$	& 0.98 \\ 
		&&&&&&&&&&& \\
		\hline
	\end{tabular}
			\begin{tablenotes}
			\item[a] * HSS where \textit{relxilllp} model is used instead of \textit{relxilllpCp}. Here, $kT_e$ is not a parameter and \textit{norm} is the \textit{relxilllp} normalization. 			
			\item[b] \dag Parameter uncertainty can't be estimated.
			\item[c] \dag \dag Frozen parameter.
			\item[d] \ddag Parameter hits the boundary.
		\end{tablenotes}  
\end{table*}

\begin{figure}
	\includegraphics[width=\columnwidth]{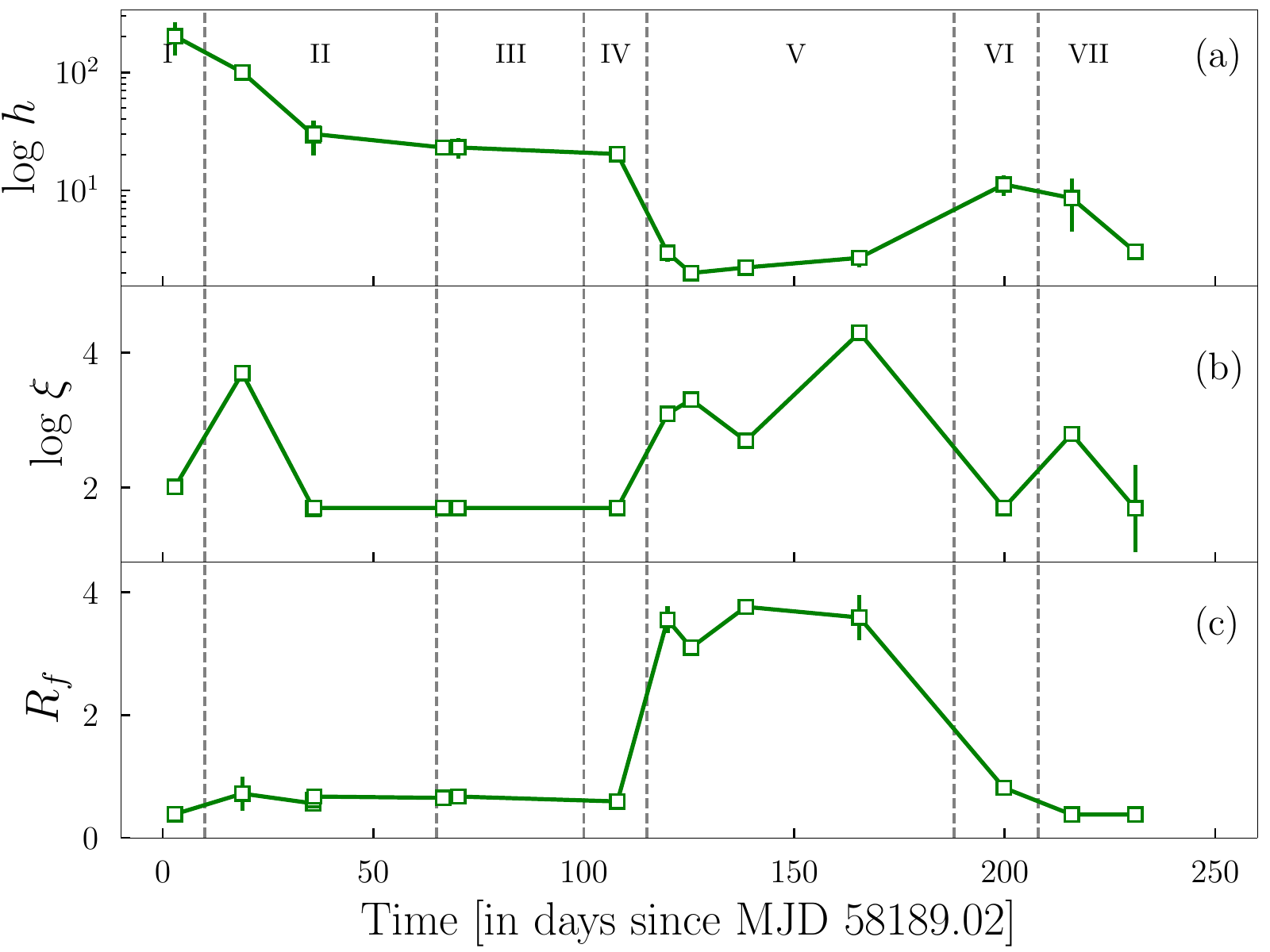}
	\caption{ Evolution of the reflection parameters: \textbf{(a)} lamp-post height (\textit{h}), \textbf{(b)} ionization parameter ($log$ $\xi$) and \textbf{(c)} reflection fraction ($R_f$) obtained from simultaneous wideband spectral study. Different evolutionary phases are marked from I to VII.}
	\label{fig:fig_refl}
\end{figure}

In the reflection modelling, we mainly use a combination of \textit{diskbb} and \textit{relxilllpCp} models to fit the data except 
Epoch $9-12$. During these four epochs, which are in the HSS, \textit{relxilllpCp} model did not produce a good fit and value of few parameters become unphysical. Therefore, we use \textit{relxilllp} for a better spectral fitting. 
We present the simultaneous NICER-NuSTAR (Epoch 1) and XRT-NuSTAR (Epoch 3) broadband reflection modelling in Fig.~\ref{fig:sim}a and Fig.~\ref{fig:sim}b respectively.
We use the model \textit{tbabs(diskbb+relxilllpCp)} to fit the data for both cases. 
Black color represents the XRT data in Fig.~\ref{fig:sim}a and NICER data in Fig.~\ref{fig:sim}b, whereas red and green are for NuSTAR FPMA and FPMB respectively. Fig.~\ref{fig:astro-refl} represents the AstroSat observation (Epoch 9) fitted with the model \textit{tbabs(diskbb+relxilllp)}.
 Here, SXT and LAXPC data are shown in magenta and green colors, respectively. As thermal disk component is needed for the low energy data. Note that in phenomenological modelling (\S\ref{sec:simu-pheno}), the reflection features are managed by a \textit{gaussian} component at $\sim$ 6.4 keV and \textit{smedge} component $\sim$ 7 keV for most of the epochs. On the other hand, reflection features such as iron line, edge and the Compton hump are self-consistently calculated in the \textit{RELXILL} model. 

Similarly, all the pairs in Table~\ref{tab:simult-obs} are fitted with RELXILL models. The important parameters are estimated and are summarised in Table~\ref{tab:simult-refl}, while other parameters are frozen at their default values. The uncertainties are within 90$\%$ confidence range. 
The inner disk temperature, $T_{in}$ and photon index, $\Gamma$ estimated for each epoch from phenomenological (Table~\ref{tab:simult-pheno}) and reflection modelling (Table~\ref{tab:simult-refl}) are generally agreed well. However, $T_{in}$ estimated from reflection modelling is marginally lower, $\Gamma$ is a little on the higher side. A possible reason could be that the soft-excess from the reflection model provides a small contribution towards lower energy reducing the temperature of the \textit{diskbb} component. Also, the primary spectrum may become marginally softer since the reflection reduces the number of hard X-ray primary photons.   

The spin parameter $a_\ast$ is found to be pegged near the highest allowed limit, therefore we freeze $a_\ast=0.998$. 
\citet{2021MNRAS.508.3104B} predicted the BH spin of MAXI J1820+070 as $0.799^{+0.016}_{-0.015}$ using timing study, but they expected an underestimation in the calculation and this value should be regarded as a lower limit. Therefore, the high spin approximation that we have used in our study is justified. The inclination angle, $\theta$, of the system is estimated to be $63^\circ$ \citep{2020MNRAS.493L..81A}. We could able to estimate the inclination angle for few observations (Epoch 3, 4, 8 and 11). Otherwise, we freeze  $\theta=63^\circ$, since we cannot constrain it from the fitting. The inner radius of the accretion disk $R_{in}$ is fixed at $R_{ISCO}$, the innermost stable circular orbit radius, and outer radius $R_{out}$ is frozen at the default value (400 $r_g$, where $r_g \equiv GM/c^2$, the gravitational radius of the BH). In most of cases the electron temperature in the corona $kT_e$ is hitting the upper boundary, therefore in such cases, we freeze it to a more reasonable value of 100 keV. 
We used lamp-post height $h$, inclination angle $\theta $, photon index of the incident radiation $\Gamma$, ionization parameter log($\xi$), iron abundance ($A_{Fe}$) of the accretion disk in terms of the solar abundance, and reflection fraction $R_f$ as the free parameters. $R_f$ is defined as the ratio of intensity of the primary photons illuminating the disk to that reach the observer \citep{2016A&A...590A..76D}. This parameter depends on the strength of the reflection component and the geometry of the system. 

The evolution of the reflection parameters is shown in Fig.~\ref{fig:fig_refl}.
The number of primary source photons illuminating the disk increases in the rising phase of the outburst and hence the ionization ($log$ $\xi $) (Fig.~\ref{fig:fig_refl}b) of the disk material increases. 
\begin{figure*}
	\centering
		\includegraphics[width=0.32\textwidth,angle =-90]{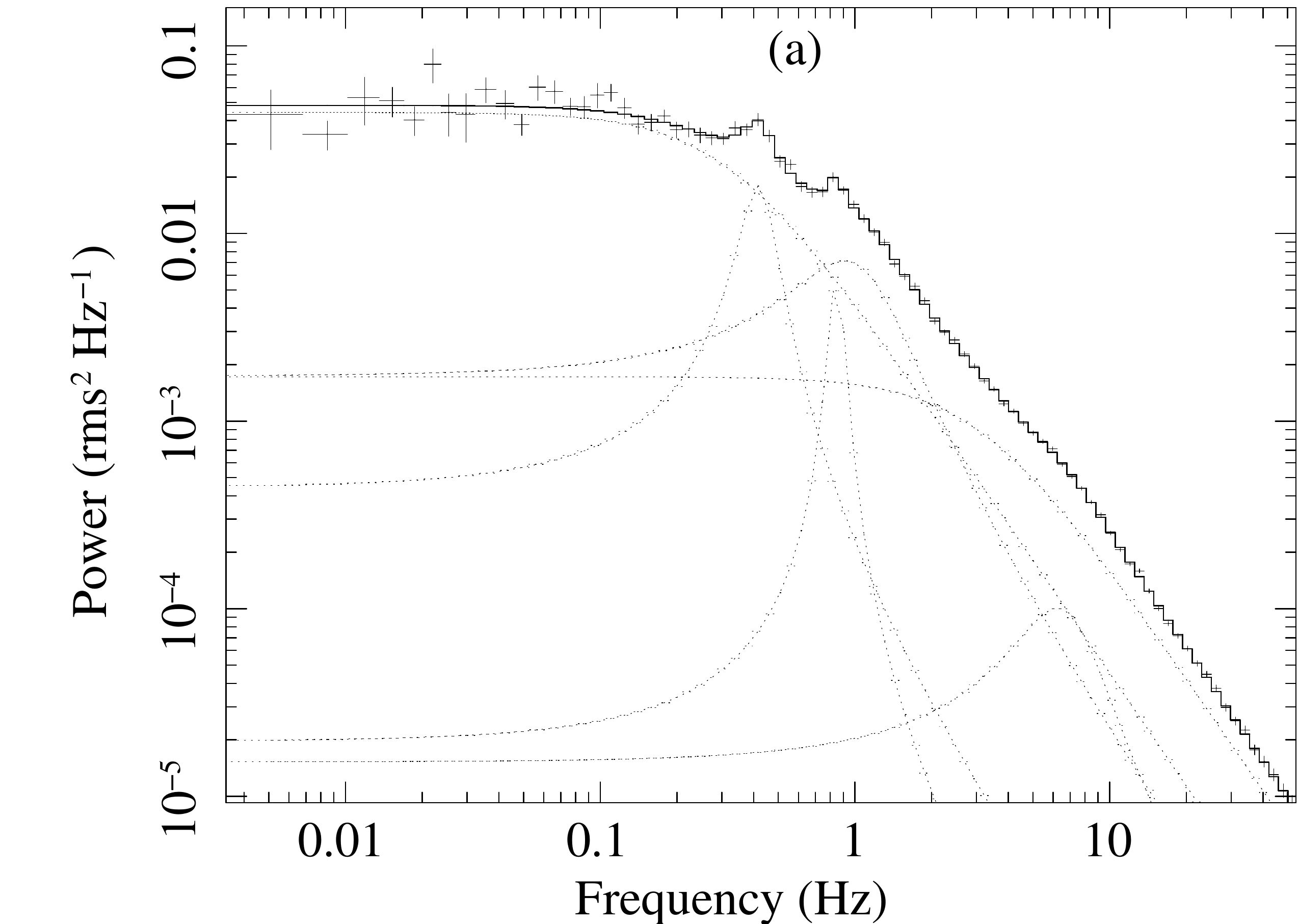}
		\includegraphics[width=0.33\textwidth,angle =-90]{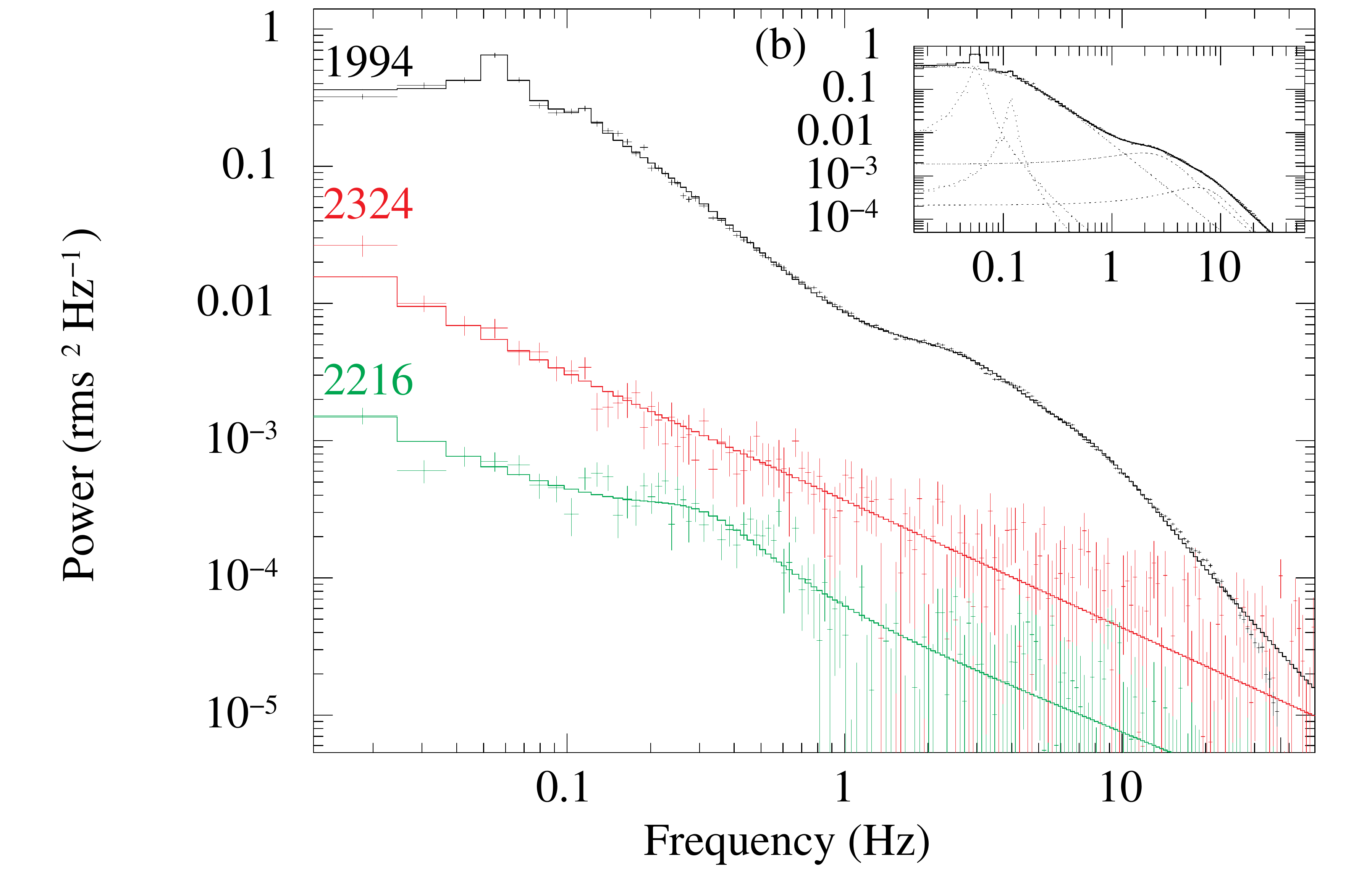}
	\caption{Representation of PDS for (a) NICER observation on day 56.43, fitted using multiple Lorentzian components and (b) AstroSat observations are shown. Three AstroSat observations on Epoch 2, 9, 12 (Table~\ref{tab:simult-obs}) are plotted in black, green, and red colors respectively. The Lorentzian components for Epoch 2 are shown in the inset.}
	\label{fig:pds-fit}
\end{figure*}
As a result, the amount of reflection, $R_f$, increases during the rising phase of Outburst-I. Therefore, $R_f$ shows a positive correlation with the luminosity of the source in the hard state \citep{2020MNRAS.493L..81A}. During Phase-II, the lamp-post height, $log$ $\xi $, and $R_f$ decrease from the rising phase whereas in Phase-III, these parameters remain steady.
$R_f$ shows a slight decrease for Epoch 8 (Phase-IV) which is the very beginning of the second outburst and has low flux. 
However, \citet{2021NatCo..12.1025Y} considered a lamp-post corona for the hard state of MAXI J1820+070 and reported an unusual decreasing trend in the reflection fraction due to the effect of the bulk motion of the corona. We are considering a point-like static lamp-post corona, where this effect is not coming into the picture. Thermal reverberation lags \citep{2021A&A...654A..14D} showed that the relative distance between the primary photon source and the disc decreases as the source softens. 
  
\citet{2016A&A...590A..76D} found that reflection is strong for low source height (h) and high inclination. 
In the HSS (Phase-V in Fig.~\ref{fig:fig_refl}), we find corona is close to the central object with a high reflection fraction $R_f \sim 3.5$, high ionization parameter ($log$ $\xi $ $\sim 4$) and over-abundance of iron ($A_{Fe}$). \citet{2019MNRAS.490.1350B} and \citet{2019MNRAS.487.5946B} also reported the chances of overestimation of $A_{Fe}$. During the decline phase (Epoch $13-15$) shows a reduction of ionization and reflection fraction. The lamp-post height shows a momentary increase (Epoch 13) when the hard flux takes over the soft flux and then declines as the source becomes faint. Thus, the broadband reflection modelling shows the presence of a dynamically evolving corona (lamp-post hight is changing) during both outbursts except in the HSS.  
\subsection{Temporal Analysis and Modelling} 
\label{sec:tempo}
For the timing analysis of MAXI J1820+070, we consider all the NICER data of the source and the three available AstroSat/LAXPC observations during this period. We generate light curves for NICER observations with more than 1 ks exposure, with a binsize of 3 ms in the full energy range. We did not subtract any background from the light curve and generated PDS using the tool \textit{powspec}. 
The PDS are Poisson noise subtracted \citep{1995ApJ...449..930Z} and normalized to squared RMS fractional variability.
For Outburst-I, we have rebinned the data such that $f_{min}=0.01$ Hz and $f_{max}=55.55$ Hz. A geometrical rebinning factor of -1.09 is applied. For Outburst-II, particularly for the soft state data, we prefer a larger binsize to get a good statistics and we have $f_{min}=0.001$ Hz and $f_{max}=2.564$ Hz. Here, we rebin the PDS with a geometrical rebinning factor of -1.2.
For AstroSat/LAXPC data, we create light curve with a time resolution of 0.1 sec in the $3-60$ keV energy band. The PDS is geometrically rebinned by a factor of -1.03. For the observation ID 9000002216 (Epoch 2, in Table \ref{tab:simult-obs}), the PDS is made only from segments 10-19 since the other segments are having low exposure time.

We generated a flat response and use XSPEC to model the PDS. The broad noise features in the PDS are modelled with zero-centered \textit{Lorentzians}. We also use powerlaw components for fitting the PDS in the HSS observations only, since the PDS are almost featureless. 
We use a combination of Lorentzian in order to identify the QPOs or QPO-like features in the PDS. The functional form of Lorentzian in XSPEC is written as, 
\begin{equation}
L(\nu)= \frac{K}{\pi} \frac{\Delta}{(\nu -\nu_c)^2+\Delta^2} \rm ,
\end{equation}
where $\nu$ is the frequency and $\nu_c$ (characteristic frequency of QPO) is the line centre. Here, $\Delta$ is the half width at half maximum (HWHM) of the line and $K$ is the normalization. All the three Lorentzian parameters ($\nu_c$, $\Delta$, and $K$) of the QPO-like features are estimated for each observation.
For each Lorentzian, we calculate Q-factor ($Q=\nu_c /2 \Delta$) and significance, which is defined as the ratio of Lorentzian normalization to its negative error \citep{2014MNRAS.445.4259A,2019MNRAS.487..928S}. The area under a Lorentzian is given by,
\begin{equation}
I= \frac{K}{2} \left[ 1+ \frac{2}{\pi}\  tan^{-1} \left( \frac{\nu_c}{\Delta} \right)  \right],
\end{equation}
and the fractional RMS amplitude is calculated as $\sqrt{I}$. The total RMS of each PDS is also computed by integrating the whole PDS over frequency \citep{2005ApJ...629..403C,2010LNP...794...53B,2011MNRAS.418.2292M,2012MNRAS.427..595M}.

Typical example of fitted PDS for NICER observations on day 56.43 is shown in Fig.~\ref{fig:pds-fit}a. Here, the PDS is fitted using multiple Lorentzian profiles.
The feature at $\nu_c$=$0.42_{-0.02}^{+0.01}$ Hz is a QPO with Q-factor $3.19\pm 1.4$, significance 2.3 and fractional RMS (in percentage) is $5.9	\pm 0.6$.
In Fig.~\ref{fig:pds-fit}b, we present PDS correspond to the AstroSat observation IDs 9000001994 (Epoch 2), 9000002216 (Epoch 9), 
and 9000002324 (Epoch 12) in black, green, and red colors respectively. Various Lorentzian components for the ID 9000001994 are shown in the inset. A QPO is detected at $0.05_{-0.001}^{+0.002}$ Hz with Q-factor of $4.1\pm 0.7$, significance of 9.1 and fractional RMS of $9.1\pm 0.3$ (in percentage) only during Epoch 2 of AstroSat observations.  
 
There are several studies in literature about the QPOs detected for the source in X-rays \citep{2019MNRAS.490.1350B,2020ApJ...889L..17M,2020ApJ...891L..29H,2020ApJ...889..142S,2021NatAs...5...94M}. 
\citet{2018ATel11578....1B} reported QPOs with evolving frequency ($0	.036_{-0.002}^{+0.002}$ to $0.428_{-0.036}^{+0.004}$ Hz) in the LHS of MAXI J1820+070.
 \citet{2020ApJ...891L..29H} reported a QPO type switching, from type-C to type-B (rapidly became 4.5 Hz then decreased to 3 Hz) during (MJD 58305.67 to MJD 58305.81) the LHS to HSS transition followed by a strong radio flare detection which provides the connection between them.
\citet{2021NatAs...5...94M} observed LFQPOs above 200 keV in the LHS using Insight-HXMT observations and such a high energy LFQPO has not been observed before in black hole binaries. They found an increase in frequency from 0.02 Hz (MJD 58194) to 0.51 Hz (MJD 58257) followed by a decrease to 0.22 Hz (MJD 58286).

Here, we have studied the QPO or QPO-like features in the PDS and tried to figure out the evolution of their characteristic frequency $\nu_c$, power under such features and their variation with $\nu_c$. Finally, we present the RID and HRD which help to track the behavior of the BH transient throughout the outburst and can provide fundamental information regarding the system.
\subsubsection{Evolution of Characteristic Frequency}
\label{sec:qpo}
\begin{figure*} 
	\centering
		\includegraphics[width=\columnwidth]{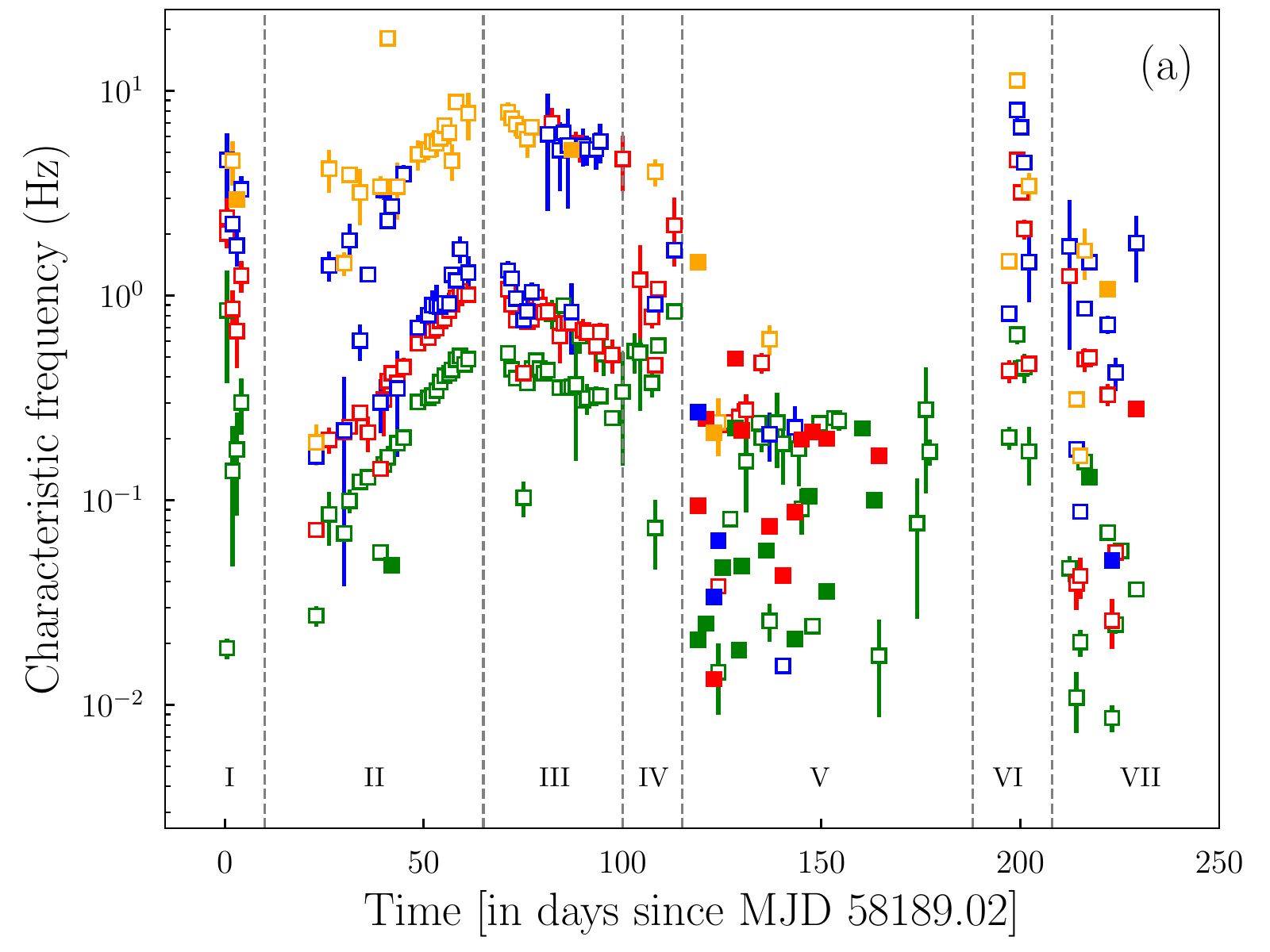}
		\includegraphics[width=\columnwidth]{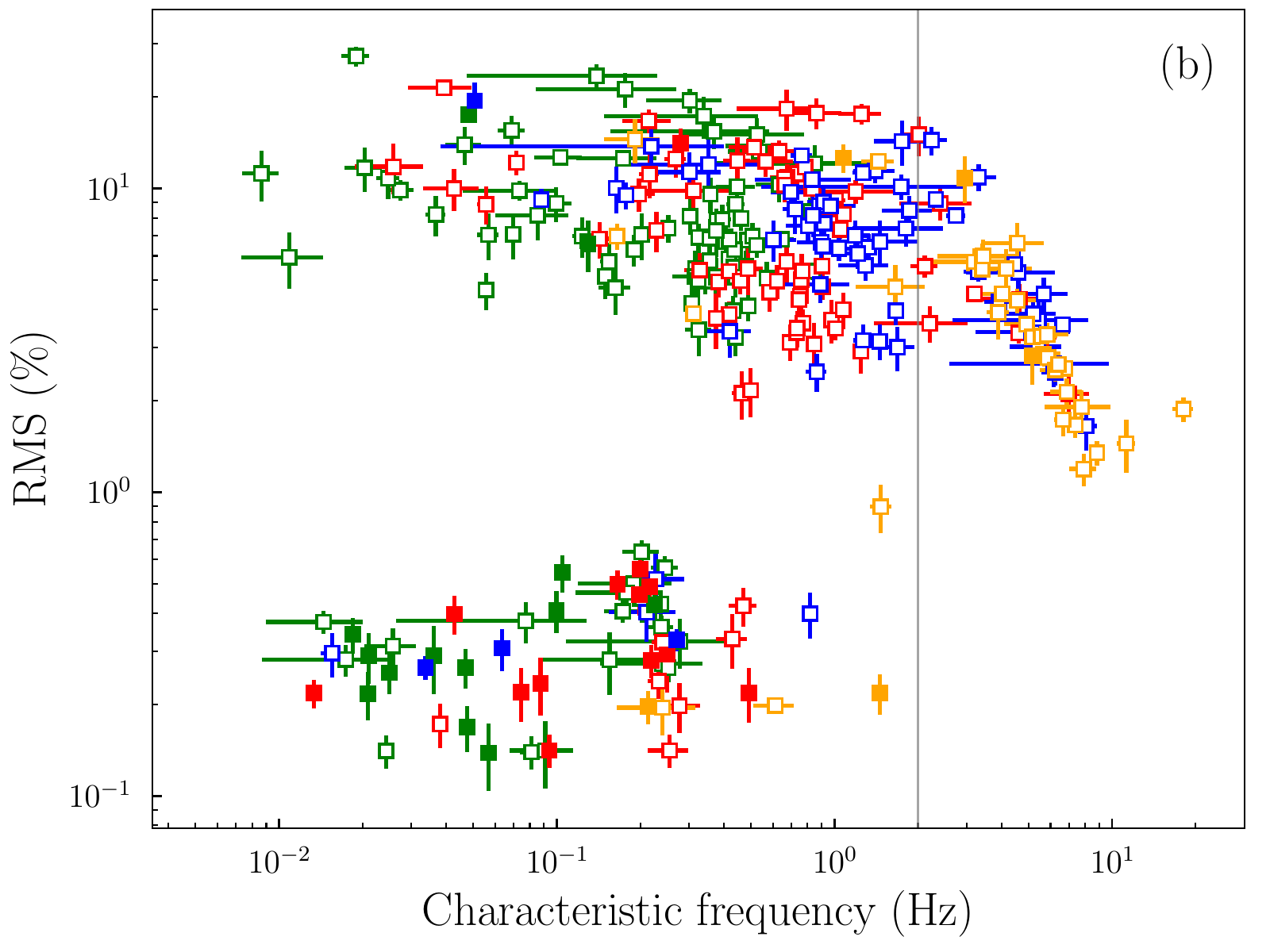}
	\caption{\textbf{(a)} Time evolution of the characteristic frequency ($\nu_c$) of different Lorentzian features corresponding to each NICER observation. First four Lorentzian of each PDS, in the increasing order of $\nu_c$, are indicated using green, red, blue and orange colors, respectively. Different evolutionary phases are marked from I to VII. \textbf{(b)} The variation of fractional RMS (in percentage) under Lorentzian with the characteristic frequency for the first four Lorentzian of NICER observations are presented with color scheme. The $\nu_c$ whose uncertainty can't be estimated are shown using filled boxes.}
	\label{fig:tim-1}
\end{figure*}

The evolution of $\nu_c$ for different \textit{Lorentzian} features (say, QPO-like features) used in each NICER observation 
is shown in Fig.~\ref{fig:tim-1}a. 
We have plotted the first four \textit{Lorentzian} of each observation in the increasing order of the value of $\nu_c$, indicated using green, red, blue, and orange colors respectively. We could not estimate the uncertainty in $\nu_c$ for a few observations, which are shown using filled squares. 
The characteristic frequency of all these \textit{Lorentzian} features increases with time in Phase-II though the overall flux decreases (Fig.~\ref{fig:lc_all}). This is contrary to the general understanding that as the source moves in the LHS, the changes in $\nu_c$ of the \textit{Lorentzian} features are correlated with the source flux \citep{2010LNP...794...53B}. However, the behavior of $\nu_c$ in Phase-III is expected as the source flux gradually declines in LHS. The presence of type-C QPO is observed in both Phase-II and Phase-III. We discuss the possible physical scenarios to understand the evolution of $\nu_c$ in \S\ref{sec:accsin}. 

In the HSS (Phase-V), the PDS is almost featureless and are fitted using weak \textit{Lorentzians} and \textit{powerlaw} components. At this state, the \textit{Lorentzians} have low $\nu_c$ compared to the previous state and no QPO is observed. During the decay towards LHS (Phase-VII) in Outburst-II, $\nu_c$ shows a similar trend as in Phase-III which is expected in LHS.

The fractional RMS (\%) under each QPO-like feature is calculated and plotted against $\nu_c$ in Fig.~\ref{fig:tim-1}b. The same color scheme (as in Fig.~\ref{fig:tim-1}a) is followed. The figure shows some interesting characteristics. Two clusterings can be distinguished here; one is at the top-right, corresponds to the LHS with higher values of RMS, and the other is at the bottom-left, representing  the HSS with very low RMS. We see that the fractional RMS shows a clear anti-correlation with QPO frequency $\nu_c$ $> 2$ Hz (marked with vertical line) which is generally observed in the LHS. However, no clear correlation is observed at the lower frequency.

\subsubsection{RMS-Intensity Diagram (RID)}
\label{sec:rid}
\begin{figure}
    \includegraphics[width=\columnwidth]{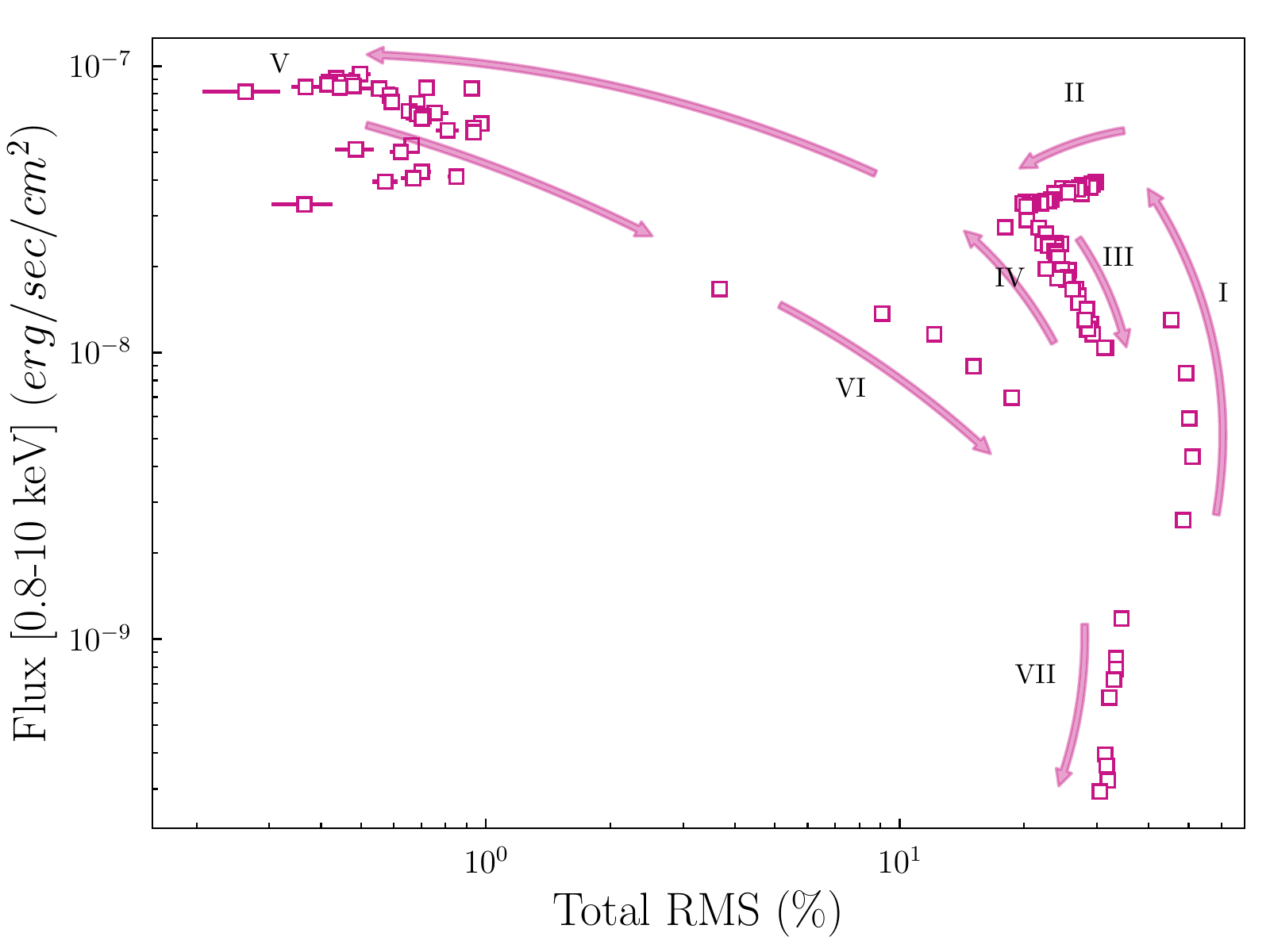}
    \caption{RID, the flux versus total RMS (in percentage) is presented with various phases identified as in Fig.~\ref{fig:lc_all} to understand the trend of evolution.}
	\label{fig:tim-2}	
\end{figure}

Fig.~\ref{fig:tim-2} shows the flux ($0.8-10$ keV) versus total fractional RMS with various evolutionary phases marked (I$-$VII). The arrow marks indicate the evolution track of the system. We can see the anti-correlation between the flux and RMS throughout this plot except in Phase-II and VII, where RMS decreases with a decrease in flux.
It shares a similar pattern as in HID (Fig.~\ref{fig:hid}).

At the beginning of the outburst (Phase-I), the flux is very low and the total fractional RMS is greater than 40\%. The flux then increases suddenly and reaches Phase-II. In this phase, RMS decreases to 20\% due to spectral softening.
But the RMS increases to $\sim$ 30\% in Phase-III since there is no spectral softening due to the reduction in thermal flux. 
The spectral state transition to HSS (Phase-V) takes place very fast and RMS drops to the lowest possible value. The flux declines gradually (Phase-V) and the source has a low RMS ($<1\%$) in this state. The flux then decreases in a faster way (Phase-VI) and reaches the quiescent state (VII) as the outburst ends. At the same time, RMS increases quickly and reaches back to the same level as in Phase-I. 
\subsubsection{Hardness-RMS Diagram (HRD)}
\label{sec:hrd}
\begin{figure}
	\includegraphics[width=\columnwidth]{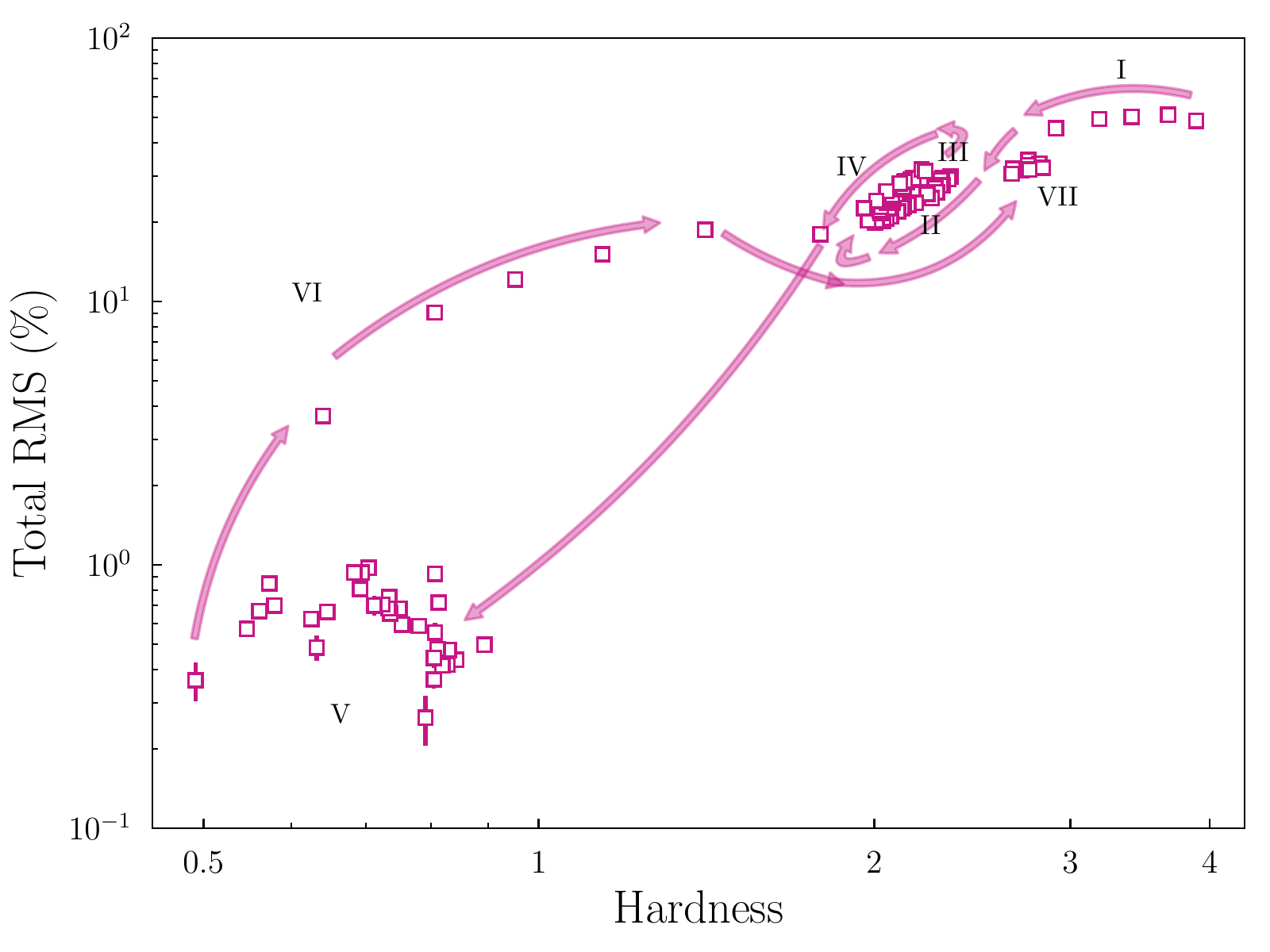}
	\caption{Hardness-RMS diagram (HRD) for the NICER observations. Total fractional RMS corresponds to the frequency range $0.02-40$ Hz in the full energy range is plotted against hardness.}
	\label{fig:tim-hrd}
\end{figure}

Total fractional RMS versus hardness (\textit{Hardness-RMS diagram}) is shown in Fig.~\ref{fig:tim-hrd}. This figure represents an overall combination of the spectral and temporal characteristics. The evolutionary track is marked from I to VII as before.

The outburst starts (Phase-I) with a high value of hardness and the total fractional RMS $>40 \%$ but as it proceeds, hardness decreases along with the RMS, and the source enters into Phase-II where RMS decreases to 20\%. Here, both are correlated due to spectral softening. Then RMS increases slightly in Phase-III with a very little change in hardness. Outburst-I ends here spanning almost a linear region with higher values of HR and RMS. During HSS (Phase-V), there is significant variation in the hardness with approximate constant ($<1\%$) RMS. Phase-VI possibly shows a state transition in the intermediate state as there are significant variations in both HR and RMS but no high frequency variability in the light curve is observed to confirm it. Finally, the source moves back to the quiescence with high HR and RMS. The Outburst-II spans a wider parameter space in HRD and completes a figure 8-shaped hysteresis loop. 

\section{A Possible Accretion Scenario}
\label{sec:accsin}
We attempt to understand the accretion scenario of MAXI J1820+070 during the 2018 outburst using standard accretion disk \citep{1973A&A....24..337S} picture. 
We find from the spectral modelling of NICER data (Fig.~\ref{fig:dbb}b) that the minimum value of \textit{diskbb norm} is,
$N_d \sim 1.5\times 10^4$ during the HSS of Outburst-II. The corresponding inner disk radius of the
\textit{diskbb} model is $r_{in}=D \sqrt{N_{d}/cos(\theta)}$, where, $D$ is the distance to the source in 10 kpc, $\theta$ is the inclination of the system in degree and $r_{in}$ is in km. Given $\theta= 63^\circ$ and $D=2.96$ kpc \citep{2020MNRAS.493L..81A}, $r_{in}=122$ km and it is consistent with the true inner radius $R_{in} \sim R_{ISCO}=0.44\, \kappa^2\, r_{in}$ \citep{1998PASJ...50..667K} of a Schwarzschild BH of mass $8.5 M_\odot$ \citep{2020ApJ...893L..37T} with $\kappa=1.7$. This shows that the accretion disk has reached close to the innermost stable circular orbit (ISCO) during the HSS of Outburst-II. Outburst-I starts with low $T_{in}\sim 0.15$ keV (Fig.~\ref{fig:dbb}a) and $N_d\sim 1.6\times 10^4$ (Fig.~\ref{fig:dbb}b). The flux reaches the peak just in 10 days. Notice that $N_d$ is very similar to the value during the HSS and hence $R_{in}$ is close to the ISCO. Therefore, both outbursts start close to the central object and are having shorter rise time.

The generally accepted disk instability model \citep{1974PASJ...26..429O,1994A&A...290..133V,2001NewAR..45..449L} suggests that an outburst can be triggered due to the thermal-viscous instability when the temperature of the disk crosses the hydrogen ionization temperature. The instability raises the temperature of the disk forming a hot propagating ionizing front which enhances the accretion of matter. Outburst can be triggered either at the outer part of the accretion disk (outside-in) or the inner part (inside-out). Since the accretion happens at the viscous timescale, the outside-in trigger will have a longer rise time ($\sim 50-100$ days) whereas inside-out trigger produces an outburst of shorter rise time ($\sim 10-20$ days). Most of the outbursts of GX 339-4 are examples of the former category \citep{2019MNRAS.486.2705A} whereas that of GRO J1655-40 and  H1743-322 belong to the second class \citep{2020A&A...637A..47A} of outbursts.

The short rise time and low $R_{in}$ is consistent with an inside-out triggering for MAXI J1820+070. Therefore, the hot ionizing front moves outward pushing the disk outward ($N_d$ increases sharply, Fig.~\ref{fig:dbb}b) with an enhancement of $T_{in}$ (Fig.~\ref{fig:dbb}a) due to a surge in accretion rate, $\dot m$, ($T_{in} \propto \dot m^{1/4} r^{-3/4}_{in}$). During the decay phase (Phase-II, III), the accretion rate continues to decline with a very slow inward movement of the disk keeping the disk temperature approximately constant. The disk is truncated in the range $13-18 r_g$ and the similar conclusion is drawn by \citet{2021arXiv211208116Z} using Insight-HXMT, NuSTAR and INTEGRAL data. 

At around $\sim$ 100 days (Phase-IV), one more inside-out triggering (Outburst-II) takes place and the next 20 days it is a repetition of the rising phase of Outburst-I. But the difference is that Outburst-I has been triggered from quiescence (low $T_{in}$) whereas the second triggering has happened when the disk is sufficiently hot ($T_{in}=0.28$ keV). Therefore, the accretion process is more dramatic, particularly in the last 5 days of the rising phase. The disk becomes so hot ($T_{in}$ increase by a factor of 2.6) that the viscous timescale reduces and the $R_{in}$ reaches (reduce by a factor of 3.5) close to $R_{ISCO}$ in just 5 days. The soft flux under the \textit{diskbb} component at the peak of Outburst-II is $5\times 10^{-8}$ erg/cm$^2$/s and it corresponds to a luminosity of $5.3\times 10^{37}$ erg/s for an assumed distance of $2.96$ kpc \citep{2020MNRAS.493L..81A}.
It means the source is accreting only at the $5\%$ of its Eddington accretion rate ($\dot M_{Edd}=$ $1.2\times 10^{19}$ gm/s) with an assumed mass of $8.5 M_\odot$ \citep{2020ApJ...893L..37T} and 10\% accretion efficiency.
Even though the source is having a low accretion rate, the system appears to be very bright due to proximity and low column density along the line of sight. 
 A low luminosity outburst is expected for an inside-out triggering due to the limited size of the disk in the hot state \citep{2001NewAR..45..449L}.
The source continues to be in HSS with approximately no change in $R_{in}$, whereas $T_{in}$ continues to decrease due to the decline of the accretion rate. Finally, the disk becomes cooler once the accretion of hot matter is over and the source reaches the quiescence.

The broadband spectral study using reflection modelling (\S\ref{sec:simu-refl}) reveals the presence of a dynamical corona in the LHS (Phase-II and Phase-III), while the inner disk remains truncated and evolves marginally. Therefore, the evolution of spectral and timing properties in this state are regulated by the dynamical evolution of the corona.
Fig.~\ref{fig:tim-1}a shows that the characteristic frequency of QPOs, $\nu_c$, increases with time in Phase-II whereas it decreases with time in Phase-III. Among the few possible physical scenarios for the origin of LFQPOs, \citet{1999ApJ...524L..63S,2018MNRAS.473..431M} and \citet{2019NewAR..8501524I}  proposed Lense-Thirring precession of inner accretion can be an effective mechanism.
Lense-Thirring precession mechanism required a misalignment of $\gtrsim 5^\circ$ between the BH spin axis and the binary orbital plane \citep{2019NewAR..8501524I,2019Natur.569..374M}. \citet{2021arXiv210907511P} recently reported that there is a misalignment of at least $40^\circ$ between the black hole spin and the orbital angular momentum for MAXI J1820+070.
Also, the inner radius of the disk needs to be dynamic to produce the QPO evolution with time. That means $R_{in}$ has to move inward and outward during Phase-II and Phase-III, respectively. However, we understand from the spectral modelling (\S\ref{sec:xrt-ni} and \S\ref{sec:simu}) that the $R_{in}$ decreases marginally ($R_{in}\sim 13-18 r_g$) during the LHS (Phase-II and III). 
\cite{2007A&ARv..15....1D} suggested that $R_{in}$ derived from \textit{diskbb norm} using simple spectral models such as $diskbb + powerlaw$ may not always be reliable. 
However, there are other studies in the literature that support the truncated disk geometry and dynamic corona in the LHS of this source. For example, \citet{2019Natur.565..198K} and \citet{2020ApJ...896...33W} used reverberation time lags between the emission from the corona and the irradiated accretion disk, and suggested a contracting corona in the LHS rather than a changing inner radius of the accretion disk. \citet{2019MNRAS.490.1350B} and \citet{2021NatCo..12.1025Y} also arrived at the similar conclusion using reflection modelling of the NuSTAR data. Similarly, \citet{2021arXiv211208116Z} performed reflection modelling of HXMT, NuSTAR and INTEGRAL data and suggested $R_{in} > 10 r_g$ in the LHS. Also, \citet{2021ApJ...909L...9Z} supported the presence of a truncated disk using thermal Comptonization and reflection modelling of NuSTAR and INTEGRAL data. Therefore, Lense-Thirring precession of inner accretion flow cannot explain the evolution of QPO frequency of the source completely.
   
However, the other competing picture for the origin of LFQPO is the propagation of shock oscillation model \citep[and reference therein]{1996ApJ...457..805M,2004A&A...421....1C}. Here, oscillation of the `hot' corona due to the resonance between the cooling timescale and the local dynamical timescale can produce LFQPOs. 
During Phase-II, the supply of soft photons remain steady as the disk is not evolving and therefore, the photon number in the corona is approximately constant (\textit{powerlaw norm} remains constant, Fig.~\ref{fig:po}b). The soft photons are going to cool the electrons in the corona and a slow but steady spectral softening takes place. It reduces the cooling timescale of electrons in the corona. If the LFQPOs are produced due to the resonance between the cooling timescale and the local dynamical timescale, the QPO frequency will increase with time. On the other hand, in Phase-III, the number of thermal photons decreases gradually which increases the cooling timescale and hence a reverse evolution is observed. 
Note that in Phase-II, the inverse-Comptonization of soft photons in the `hot' corona is significantly high as the soft photon source remains steady. The hard photons lag behind the soft photons due to inverse-Comptonization effect. The reflection modelling (Fig. \ref{fig:fig_refl}a) shows that the corona is dynamic and moving closer to the central source in Phase-II. It means that the size of the corona becomes smaller and optical depth decreases in this phase. In an optically thin medium, the inverse-Compton spectral index is $\alpha=-$ln $\tau/$ ln $A $ \citep{1979rpa..book.....R}, where $\tau$ is the optical depth and $A$ is the mean amplification factor per scattering. The decrease of optical depth or/and reduction of amplification factor due to cooling can cause spectral softening observed in Phase-II. In the former case, considering $A$ to be constant, the number of scatterings required to produce a high energy photon remain constant, and the soft photon stays longer inside the corona due to an increase in the mean free path. In the second case, assuming a constant mean free path, the number of scatterings required to produce a high energy photon increase as $A$ reduces, and the soft photon spends a longer time inside the corona. The physical situation is possibly a combination of both effects mentioned above. Therefore, spectral softening is associated with the increase of hard lag in Phase-II, as \citet{2020ApJ...896...33W} reported. However, in Phase-III, the soft photons flux decreases gradually, spectra become slightly harder due to a reduction of cooling and the hard lag decreases \citep{2020ApJ...896...33W}.
Also, we have seen in Phase-II, the size of the corona decreases (Fig. \ref{fig:fig_refl}a) as the local dynamical timescale decreases and a reverse trend is expected in Phase-III. 
However, two closely spaced simultaneous data points in phase-III (Fig. ~\ref{fig:fig_refl}a) are not sufficient to comment if corona moves outward.

The geometry of the corona is generally described either as vertically elongated \citep{1996MNRAS.282L..53M,2014A&ARv..22...72U,2016ApJ...821L...1N} or radially extended \citep{1995ApJ...455..623C,2015MNRAS.448..703W,2018A&A...614A..79P}. 
We understand that the reflection modelling of the broadband spectral data supports a vertically elongated corona. In contrast, the evolution of LFQPO caused by the propagation of the shock oscillation model suggests a radially extended corona inside the truncated disk.  
The comprehensive spectral and timing studies suggest a complex geometry of the corona \citep{2015ApJ...808..122F,2016ApJ...821L...1N,2016MNRAS.458..200W,2017MNRAS.472.1932G,2020ApJ...893...97Z} in the LHS of the source. 
It demands further investigation of the coronal geometry, which is beyond the scope of the present work.

\section{Summary and Conclusion}
\label{sec:concl}
We have performed spectral analysis of NICER and Swift/XRT data (\S\ref{sec:xrt-ni}) independently during Outburst-I and Outburst-II.  
The NICER and XRT light curve is shown in Fig. \ref{fig:lc_all}, which has two successive outbursts during this period. The Outburst-I is dominated by the non-thermal photons (peak flux in $2-10$ keV is 2.4 times the same in $0.8-2$ keV) whereas the Outburst-II is very bright and soft. The same trend is found in $10-60$ keV (filled star in Fig.~\ref{fig:lc_all}) from simultaneous broadband spectral modelling. The HID with different spectral states is presented in 
Fig.~\ref{fig:hid}. The Outburst-I is evolved only in the LHS and classified as a `failed' outburst, whereas the Outburst-II shows a transition from LHS to HSS via a very brief intermediate state. There is an overall offset ($\sim$ 22\%) in the HR between XRT and NICER due to difference in the spectral responses (discussed in Appendix-\ref{sec:appxa}). 
We have done broadband spectral modelling of all available simultaneous data of NICER-NuSTAR, XRT-NuSTAR and AstroSat (SXT-LAXPC) to understand the presence of various model components (\S\ref{sec:simu-pheno}) in the spectra. Also, the broadband reflection modelling (\S\ref{sec:simu-refl}) shows the presence of strong reflection signature during both outbursts and presence of an evolving corona in the LHS. The timing analysis shows the presence of multiple QPO-like features and we have studied the time evolution of their characteristic frequency 
(Fig.~\ref{fig:tim-1}a). The RMS under the \textit{Lorentzian features} in the PDS indicates an interesting correlation. 
We find a clear anti-correlation between the fractional RMS and $\nu_c$ (Fig.~\ref{fig:tim-1}b) for $\nu_c>2$ Hz. 
This is a signature of type-C QPO and is commonly observed in LHS \citep{2010LNP...794...53B,2022MNRAS.510.3019A}. 
However, no clear correlation is observed at the lower frequency.
Below, we summarise the important findings of this work combining the spectral and timing studies during various phases of evolution of the source.
\begin{itemize}
\item \textbf{Phase-I (Rising phase, Outburst-I)}\\
Fig.~\ref{fig:dbb} and Fig.~\ref{fig:po} show that both thermal and non-thermal flux increases in the rising phase with an increase of $T_{in}$ and \textit{powerlaw norm} with a hard spectrum ($\Gamma \sim 1.5$). The NICER and NuSTAR data show the presence of Fe $K_\alpha$ line (Fig.~\ref{fig:gauss} and Epoch 1 in Table \ref{tab:simult-pheno}) and the presence of significant reflection. The lamp-post height is significantly large ($\sim200 r_g$) with a high ionization parameter ($log$ $\xi=2.01^{+0.08}_{-0.06}$) and iron abundance $A_{Fe}=2.4 \pm 0.1$. PDS shows the presence of QPO-like features in $0.1-5$ Hz (Fig.~\ref{fig:tim-1}a) and the fractional RMS $\sim 40\%$ (Fig~\ref{fig:tim-2}).\\

\item \textbf{Phase-II}\\
In this phase, the thermal disk component is very steady with $T_{in} \sim 0.27$ keV (Fig.~\ref{fig:dbb}a) and $R_{in}$ evolves very little ($15-18 r_g$). The non-thermal photon flux also remains steady with the gradual spectral softening (Fig.~\ref{fig:po}a). 
We observe reflection signature of Fe $K_\alpha$ line (Fig.~\ref{fig:gauss}) with an increasing trend in the line energy and absorption edge (Epoch $2-5$ in Table \ref{tab:simult-pheno}). The reflection modelling (Epoch $2-5$ in Table \ref{tab:simult-refl}) of broadband data shows a gradual decrease of lamp-post height to $\sim 30 r_g$ with an enhancement in the reflection fraction $R_f \sim 0.6$. The disk iron abundance remains similar to the rising phase but there is a reduction in the ionization parameter ($log$ $\xi\sim 1.7$). The timing results show that the characteristic frequency ($\nu_c$) increases with time from $\sim 0.1$ to $10$ Hz (Fig.~\ref{fig:tim-1}a) as the flux decreases.
This is generally not observed in outbursting sources. As we have discussed in \S\ref{sec:accsin}, one possible reason could be the spectral softening due to the reduction of cooling timescale of electrons in the corona as the thermal component remains steady. And the resonance oscillation between cooling timescale and dynamical timescale produces the LFQPOs \citep{1996ApJ...457..805M,2004A&A...421....1C}. The spectral softening leads to a correlated reduction of the fractional RMS (Phase-II in Fig. \ref{fig:tim-2}) with the observed flux. The RMS lies in $30-20\%$ range with a reduction in HR in this phase (Fig. \ref{fig:tim-hrd}). \\         

\item \textbf{Phase-III}\\
A steady decrease in the thermal and non-thermal components of flux is observed in this phase of the outburst. 
A steady decrease in $T_{in}$ (Fig.~\ref{fig:dbb}a) and \textit{powerlaw norm} (Fig.~\ref{fig:po}b) is observed  without any spectral softening ($\Gamma \sim 1.7$).
The NICER data possess reflection signature of Fe $K_\alpha$ line (Fig.~\ref{fig:gauss}).
The wideband spectral study shows the presence of Fe absorption edge (Epoch 6-7 in Table \ref{tab:simult-pheno}). The reflection parameters (Fig. \ref{fig:fig_refl}) are  similar to the previous phase.
We observed that the characteristic frequencies ($\nu_c$) in the PDS decreases (Fig.~\ref{fig:tim-1}a) with time. The total fractional RMS increases with the decrease in flux (Phase-III in Fig.~\ref{fig:tim-2}) which is consistent with the general characteristic in the LHS. This is due to the overall decrease in the number of photons in corona and it becomes difficult to cool the corona further. The RMS increases to 30\% with almost constant HR (Fig. \ref{fig:tim-hrd}) since both thermal and non-thermal flux decrease in a similar manner. 

After this, the source moves to Outburst-II through a rising phase (Phase-IV) similar to the Outburst-I. The simultaneous broadband data show reflection signature (Fig.~\ref{fig:fig_refl}) with similar parameter as in Phase-II. The characteristic frequency, $\nu_c$, increases with time (Fig.~\ref{fig:tim-1}a) as flux increases. Both fractional RMS and HR reduce and the source moves to HSS very fast.\\

\item \textbf{Phase-V (HSS, Outburst-II)}\\
The spectrum in this state is dominated by the thermal emission flux. The disk moves close to the inner edge with $T_{in} \sim 0.7$ keV (Fig.~\ref{fig:dbb}a) and spectrum becomes very soft, $\Gamma > 3.0$ (Fig.~\ref{fig:po}a), though simultaneous spectral modelling (Epoch 9-12) in Table \ref{tab:simult-pheno} and Table \ref{tab:simult-refl} provide relatively lower index. The NICER data do not show any Fe line signature though simultaneous broadband data show high reflection fraction $R_f \sim 3.5$, high ionization parameter ($log$ $\xi\sim 4$), and over-abundance of iron. The timing properties show featureless PDS with very low RMS (Phase-V in Fig. \ref{fig:tim-2}) and HR varies between $\sim 0.5-0.7$ (Fig. \ref{fig:tim-hrd}).\\
   
\item \textbf{Phase-VI \& VII (Decay towards quiescence)}\\
The light curve in the decline phase of Outburst-II is initially linear and becomes exponential towards the end. $T_{in}$ decreases (Fig.~\ref{fig:dbb}a), $R_{in}$ gradually move outward and the spectrum becomes harder (Fig.~\ref{fig:po}a). The reflection modelling (Fig.~\ref{fig:fig_refl}) shows that the reflection fraction and the lamp-post height decrease gradually. 
Fig.~\ref{fig:tim-1}a shows that $\nu_c$ increases at the end of HSS and decreases with the declination of the flux. The spectro-temporal properties indicate that this phase passes through the intermediate state (Phase-VI in Fig. \ref{fig:tim-hrd}) with a fractional RMS of 10\% and HR $\sim 1$. At the end, the source returns back to LHS (Phase- VII in Fig. \ref{fig:tim-hrd}) with RMS $\sim$ 30\% and HR $\sim 3$. 

\end{itemize}

Finally, we discuss the possible accretion scenario during the 2018 outburst in \S\ref{sec:accsin}. We propose that both outbursts are triggered as inside-out propagation of thermo-viscous instability which produces outbursts with a short rise time. The spectral modelling shows that the Outburst-I has been triggered close to the inner edge and the disk remains truncated during Outburst-I. The rise of flux in Outburst-II is dramatic since the outburst triggering has happened when the disc is sufficiently hot and the viscous timescale becomes short enough that the inner disk moves close to ISCO very fast and produce bright HSS. The accretion disk remains truncated ($\sim 10$ $r_g$) once the source reaches the quiescence. 
The comprehensive spectral and timing studies reveal that a lamp-post corona or a radially extended corona is inadequate to interpret the observed data. The results suggest a complex corona geometry in the LHS of MAXI J1820+070. The outcome of this study challenges the present understanding of coronal geometry and suggests further investigation of disk-jet geometry around black holes.

\section*{Acknowledgements}
We thank the reviewer for his/her valuable comments and suggestions that helped us to improve the quality of the paper significantly.
This publication uses data from the NICER and NuSTAR mission by the National Aeronautics and Space Administration. This work made use of data supplied by UK Swift Science Data Centre.
Also this work made use of data from the AstroSat mission of the ISRO archived at the Indian Space
Science Data Centre (ISSDC). This work has been performed utilizing the calibration databases and auxiliary analysis tools developed, maintained and distributed by AstroSat-SXT team with members from various institutions in India and abroad. Also this research made use of software provided by the High Energy Astrophysics Science Archive Research Center (HEASARC) and NASA's Astrophysics Data System Bibliographic Services. GP thanks H. Sreehari, Postdoctoral Fellow, IIA, India for his support toward data analysis. AN thanks GH, SAG, DD, PDMSA and Director, URSC for encouragement and continuous support to carry out this research. AMP thanks RESPOND (Research Sponsored) program Sanction order No. DS-2B-13012(2)/19/2019- Sec.II for the financial support. \\
Facilities: Swift, NICER, NuSTAR, AstroSat

\section*{Data Availability}

The data from Swift/XRT, NICER and NuSTAR underlying this article are available in HEASARC, at \url{https://heasarc.gsfc.nasa.gov/docs/archive.html}.
AstroSat data archive is available at \url{https://astrobrowse.issdc.gov.in/astro_archive/archive/Home.jsp}.


\bibliographystyle{mnras}
\bibliography{geethu} 

\appendix
\section{XRT and NICER Spectral Discrepancies}
\label{sec:appxa}
\begin{figure*}
	\centering
		\includegraphics[width=0.30\textwidth,angle =-90]{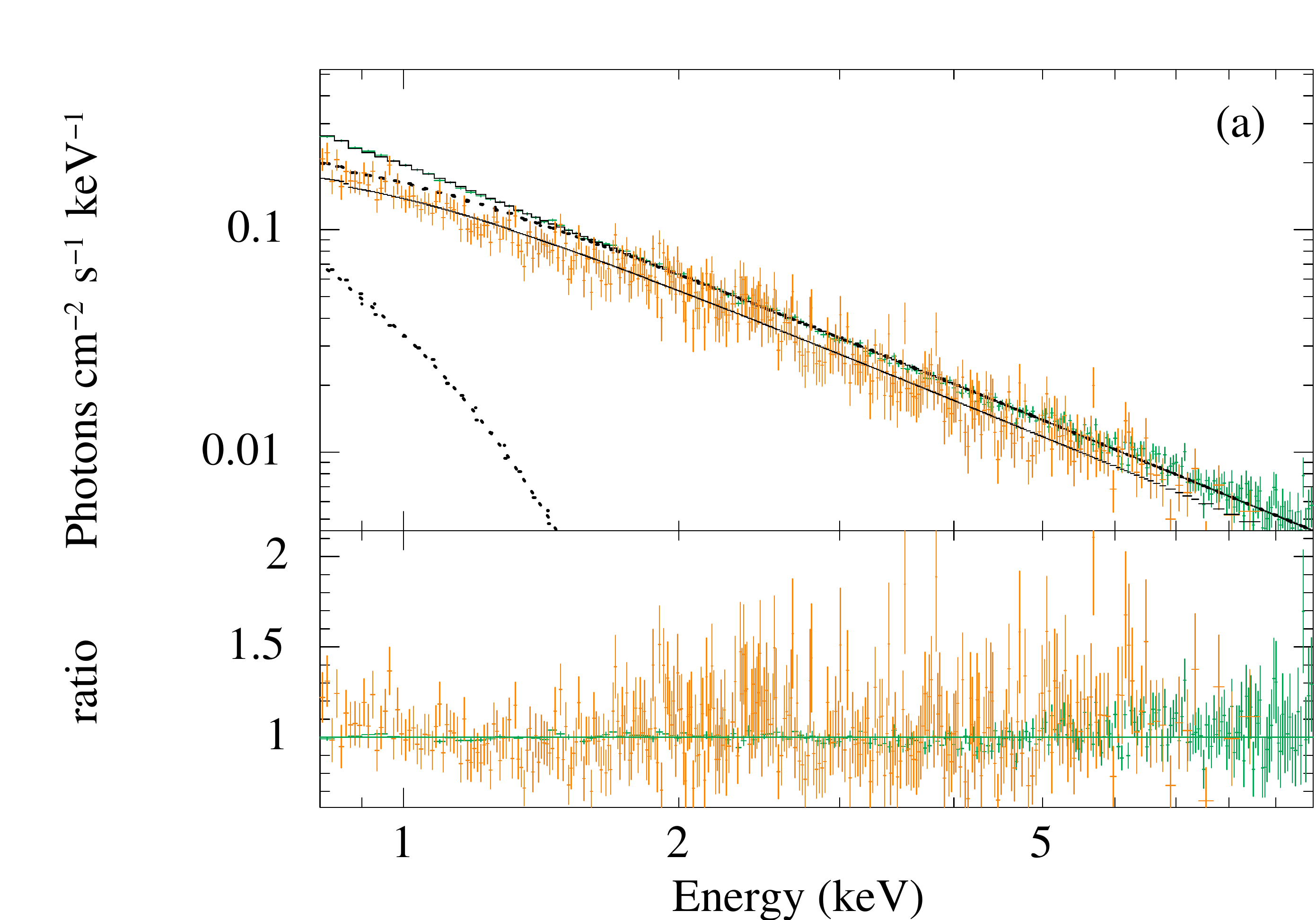}
		\includegraphics[width=0.30\textwidth,angle =-90]{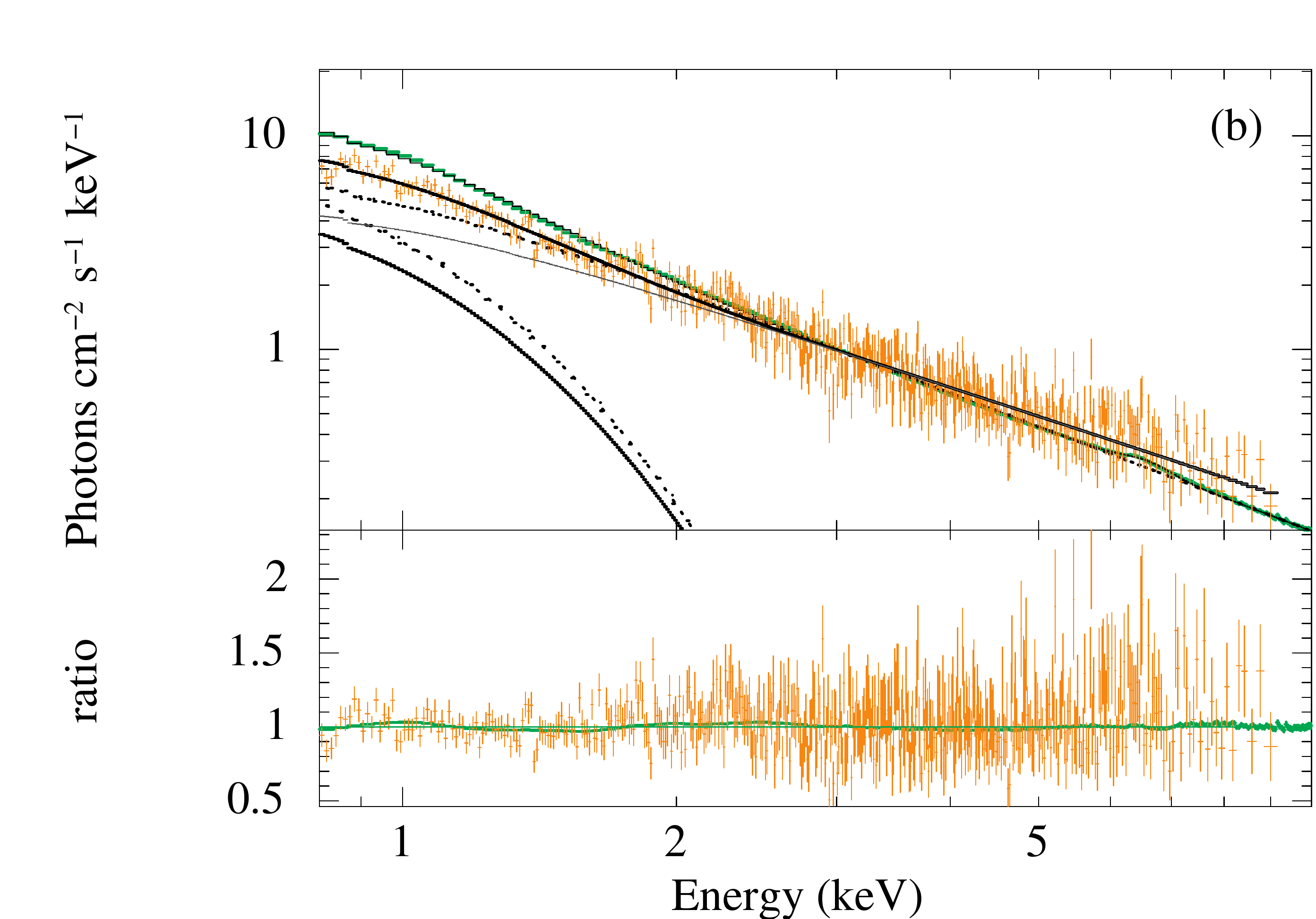}
	\caption{\textbf{(a)} Simultaneous XRT and NICER spectrum on MJD 58212 fitted using \textit{tbabs(diskbb+po)} for XRT and \textit{tbabs(diskbb+gauss+po)} for NICER, and are plotted together. The orange color corresponds to the XRT and green is for NICER. \textbf{(b)} Quasi-simultaneous observation on $\sim$ MJD 58401 are fitted using \textit{tbabs*po} and \textit{tbabs(diskbb+po)} for XRT and NICER respectively with the same color scheme.}
	\label{fig:appxfig}
\end{figure*}
The estimation of spectral parameters gets affected due to the difference in spectral energy resolution of XRT and NICER. 
To understand it better, we model two pairs of simultaneous NICER and XRT observations, the first one is on MJD 58212 (Pair-I) 
and the second observation (Pair-II) is on $\sim$ MJD 58401 (Pair-II is a quasi-simultaneous, but the source did not evolve significantly during the observations).

Pair-I observation is close to the peak of Outburst-I and the model \textit{tbabs(diskbb+po)} provides a good fitting for XRT data whereas NICER data need an additional Gaussian component due to Fe $K_\alpha$ emission line signature. The fitted spectra for both XRT and NICER data are plotted together in Fig.~\ref{fig:appxfig}a. Here, the orange color represents the XRT spectrum and green color represents NICER spectrum. Pair-II observation is close to the quiescence after Outburst-II. The \textit{tbabs(diskbb+po)} model is needed to fit the NICER data but only a \textit{powerlaw} component (\textit{tbabs $\times$ po}) is enough to fit the XRT data and the model fitting is presented in Fig.~\ref{fig:appxfig}b.
For a given simultaneous observation, the requirement of additional model components for NICER data imply that the NICER calibration is still updating since Swift/XRT is well calibrated with other instruments \citep{2017AJ....153....2M}.

We attempt an order of magnitude estimation of the spectral discrepancies between the instruments from these simultaneous observations.
The model fitted parameters and estimated fluxes in various energy bands are summarised in Table~\ref{tab:appx-1}.  
We find that the overall ($0.8-10$ keV) flux matches well when the source was bright (Pair-I) though NICER flux in the high energy ($2-10$ keV) band is 7.1\% lower than the XRT flux due to a steeper spectral index (difference of $\sim$ 0.24). However, the inner disk temperature, $T_{in}$ is comparable but a higher \textit{diskbb norm} reflects 5.8\% higher flux in low energy ($0.8-2$ keV) band. Therefore, there is a mutual balance between the low and high energy fluxes, keeping the overall fluxes comparable when the source is bright. Note that the XRT flux in $0.8-10$ keV band is used as a reference to calculate the percentage of flux difference. 
The spectral index for Pair-II data set appears comparable. But a higher \textit{powerlaw norm} and the presence of additional \textit{disk} components record 13.6\% and 7.3\% higher flux in the high and low energy band respectively for NICER. Therefore, we find considerable amount of discrepancies between fluxes in the high and low energy bands due to the differences in spectral response. 

Finally, the simultaneous multi-mission observation campaign of quasar 3C 273 involving
NuSTAR, Swift, and NICER were performed on 2019-07-02, 2020-07-06, and 2021-06-09. The recent report \citep{2021arXiv211101613M} of International Astronomical Consortium for High Energy Calibration (IACHEC) has mentioned that the data analysis of these observations is underway. Therefore, the NICER calibration is still evolving and a proper NICER-Swift/XRT cross-calibration is very much required.

\begin{table*}
	\caption{Model fitted parameters and estimated fluxes in various
energy bands from simultaneous NICER-XRT fittings}
	\label{tab:appx-1}
	\begin{tabular}{|c|c|c|c|c|c|} 
		\hline
		\multirow{2}{*}{Model} & \multirow{2}{*}{Component}  & \multicolumn{2}{c|}{Pair I}  & \multicolumn{2}{c|}{Pair II}\\
		\cline{3-6} 
		& & NICER  & XRT & NICER & XRT \\
		\hline
		&&&&& \\
		\textit{powerlaw} & $\Gamma$ & $1.648_{-0.004}^{+0.004}$ & $1.41_{-0.07}^{+0.07}$ & $1.70_{-0.01}^{+0.01}$ & $1.69_{-0.03}^{+0.03}$ \\
		&&&&& \\
		& \textit{norm} & $6.16_{-0.04}^{+0.04}$ & $4.68_{-0.05}^{+0.05}$ & $0.214_{-0.003}^{+0.003}$ & $0.179_{-0.004}^{+0.004}$ \\
		&&&&& \\
		\textit{diskbb} & $T_{in}(keV)$ & $0.283_{-0.004}^{+0.004}$ & $0.30_{-0.03}^{+0.04}$ & $0.18_{-0.01}^{+0.01}$ &  - \\
		&&&&& \\
		& \textit{norm} ($\times 10^4$)& $11.1_{-0.63}^{+0.68}$ & $6.78_{-2.4}^{+4.7}$ & $1.43_{-0.47}^{+0.78}$ & - \\
		&&&&& \\
		\textit{gauss} & \textit{line E} (keV) & $6.60_{-0.04}^{+0.04}$ & - & - & -\\
		&&&&& \\
		& \textit{sigma} (keV) & 0.6$^\dagger$ & - & - & -\\
		&&&&& \\
		& \textit{norm} & $0.034_{-0.002}^{+0.002}$ & -& - & - \\
		&&&&& \\
		$ \chi^2_{red}$ & & 1.02 & 1.16 & 1.04 & 0.86  \\
		&&&&& \\
		\hline
		\multirow{8}{*}{Flux (ergs/sec/cm$^2$)} & Energy range & $(\times 10^{-8})$ & $(\times 10^{-8})$ & $(\times 10^{-9})$ & $(\times 10^{-9})$ \\ 
		\cline{2-6} 
		 &&&&& \\ 
		 & 0.8-10 keV & $3.945_{-0.002}^{+0.002}$ & $4.00_{-0.06}^{+0.03}$ & $1.189_{-0.007}^{+0.004}$ & $0.98_{-0.01}^{+0.01}$ \\
		 &&&&& \\
		& 0.8-2 keV & $1.180_{-0.002}^{+0.001}$ & $0.948_{-0.020}^{+0.003}$ & $0.314_{-0.003}^{+0.007}$& $0.242_{-0.003}^{+0.002}$ \\
		&&&&& \\
		& 2-10 keV & $2.765_{-0.002}^{+0.002}$ & $3.05_{-0.04}^{+0.06}$ & $0.874_{-0.005}^{+0.005}$ & $0.74_{-0.01}^{+0.01}$ \\
		&&&&& \\
		\hline
	\end{tabular}
	    \begin{tablenotes}[flushleft]
          \item $\dagger$ frozen parameter.
        \end{tablenotes}  
\end{table*}


\bsp	
\label{lastpage}
\end{document}